\documentclass[a4paper,11pt]{article}
\pdfoutput=1 
\usepackage{jheppub} 
\usepackage{amssymb}
\usepackage{amsmath}
\usepackage{amsmath,amssymb}
\usepackage{graphicx}
\usepackage{subcaption}
\usepackage{makecell}

\def\co{\mathcal{O}}

\title{Machine Learning Etudes in Conformal Field Theories}

\author[a]{Heng-Yu Chen,}
\author[b,c,d]{Yang-Hui He,}
\author[e]{Shailesh Lal,}
\author[f]{M. Zaid Zaz}
\affiliation[a]{Department of Physics,National Taiwan University, Taipei 10617, Taiwan,}
\affiliation[b]{Department of Mathematics, City, University of London EC1V 0HB UK,}
\affiliation[c]{Merton College, University of Oxford, OX1 4AW, UK,}
\affiliation[d]{School of Physics, NanKai University, Tianjin, 300071, P.R. China.}
\affiliation[e]{Faculdade de Ciencias, Universidade do Porto,\\687 Rua do Campo Alegre, Porto 4169-007, Portugal.}
\affiliation[f]{Department of Astronomy and Astrophysics, Tata Institute of Fundamental Research,
\\Mumbai, 400005, India.}

\abstract{We demonstrate that various aspects of Conformal Field Theory are amenable to machine learning. 
Relatively modest feed-forward neural networks are able to distinguish between scale and conformal invariance of a
three-point function and identify a crossing-symmetric four-point function
to nearly a hundred percent accuracy. Furthermore, 
neural networks are also
able to identify conformal blocks appearing in a putative CFT four-point function and predict the values of
the corresponding OPE coefficients. 
Neural networks also successfully classify primary operators by their 
quantum numbers under discrete symmetries in the CFT from examining OPE data. 
We also demonstrate that neural networks are able to learn the available OPE data for scalar correlation function in the 3d Ising model and predict the twists
of higher-spin operators that appear in scalar OPE channels
by regression.}

\emailAdd{heng.yu.chen@phys.ntu.edu.tw}
\emailAdd{hey@maths.ox.ac.uk}
\emailAdd{sl@fc.up.pt}
\emailAdd{mohammadzaz@gmail.com}

\begin{document} 
\maketitle
\flushbottom

\section{Introduction and Summary}\label{Sec:Intro}
\paragraph{}
Conformal Field Theories (CFTs) are a particularly interesting and important class of
Quantum Field Theories, both from a theoretical and a phenomenological
point of view. For example, as is well known, they are the natural
language to describe
physical systems undergoing phase transitions. Furthermore, in two space-time
dimensions, they admit a great deal of analytic control due to the 
presence of an infinite dimensional Virasoro symmetry. In higher than two 
dimensions, there have also been been significant advances ranging from the discovery of 
new bose-fermi symmetries in three 
dimensions \cite{Polyakov:1988md,Shaji:1990is,Fradkin:1994tt,Giombi:2011kc,Aharony:2011jz,Aharony:2012nh}
, duality to higher-spin theories in Anti-de Sitter spaces
\cite{Klebanov:2002ja,Sezgin:2002rt,Gaberdiel:2010pz,Beccaria:2014zma,Bae:2016hfy}, 
as well as a large amount
of work in computing the physical data (i.e. spectrum, OPE coefficients)
of a given field theory both numerically \cite{Rychkov:2011et,ElShowk:2012ht,El-Showk:2014dwa,Kos:2016ysd,
Simmons-Duffin:2016wlq,Chester:2019ifh,Rong:2017cow,Stergiou:2018gjj},
and analytically \cite{Beem:2014zpa,Beem:2015aoa,Alday:2016njk,Gopakumar:2016cpb,Alday:2017zzv,Aharony:2018npf}.
\paragraph{}
An important driver in this progress has been the theoretical and numerical advancements in the conformal bootstrap in
greater than two dimensions. Hence
we now have the tools to access 
a large amount of precise physical data for CFTs in various dimensions, 
along with their rich theoretical structures. With the further advances in 
computational ability, and the conceptual progress in conformal bootstrap methods, one may optimistically expect these developments will usher the usage of machine-learning (ML) techniques in the studies of conformal field theories.
Such an approach is very much in line with the programme of \cite{He:2017aed,He:2017set} to machine-learn structures arising in theoretical physics and pure mathematics {\it without} the neural network been given any prior knowledge of the underlying mathematics.
Indeed, \cite{He:2017aed,He:2017set,Krefl:2017yox,Ruehle:2017mzq,Carifio:2017bov,Erbin:2018csv} 
brought ML to string
theory.
The reader is also referred to \cite{Hashimoto:2018ftp} for the fascinating idea that space-time itself might be a neural network, as well as the introductory books \cite{He:2018jtw,RUEHLE20201} (the monograph \cite{RUEHLE20201} is an excellent technical introduction of ML to theoretical physicists). Further interesting 
applications of ML to string theory,
supergravity and quantum field theory are
\cite{Comsa:2019rcz,Krishnan:2020sfg,Chernodub:2019kon,Chernodub:2020nip}.
\paragraph{}
In this paper, we shall take some initial steps in applying the methods from machine learning to analyze the data already available to us. The aim is to build up a toolbox which we hope will become useful in tackling wide range of interesting physical problems in conformal field theories.
We will demonstrate that several definitive features of conformal field
theories, from their general structures to the physical data for the 
systems of particular importance are indeed machine learnable.

\paragraph{}
We begin with a warm-up subsection \ref{Subsec:Discrete}, where we demonstrate that machine is capable of learning to classify even and odd functions, as well as functions with antipodal  symmetries.
In subsection \ref{Subsec:Scale-Conformal}, we will show that machine learning algorithms perform well on recognizing the presence of conformal symmetry in a putative system as they can
distinguish the scale and conformal invariant {two} and three-point functions of scalar operators to near hundred percent accuracy. 
As shown in subsection \ref{Subsec:Crossing-4pt}, these algorithms are also able to recognize the crossing symmetric four-point functions again with a near hundred percent accuracy.
\paragraph{}

We next turn to Section \ref{Section:Conformal Expansion} where we focus on extracting more fine-grained 
information about conformal symmetry as carried by a putative four-point function. In particular, we 
demonstrate in a simple context that
neural networks can predict the value of the particular OPE coefficient appearing in front of a conformal
block to reasonable accuracy. As a warm-up to this, we consider the same problem, but for the Fourier
expansion of an arbitrary function. We also find along the way that neural networks can easily identify classes of functions
where a particular conformal block or Fourier mode is missing in the expansion. This somewhat curious and unexpected
property however can be understood from a Principal Component analysis, as we also show.
This serves as the first step towards applying machine-learning techniques to tackle fully-fledged conformal bootstrap
problems.

\paragraph{}
Furthermore, we show in subsection \ref{Subsection:Z2-Ising}, that these algorithms also perform well
on the occurrence of discrete symmetries in the theory. They are able to,
for instance, organize the (partial) spectrum of the 3d Ising model into
$\mathbb{Z}_2$ even and odd operators from the knowledge of their OPE coefficients. 
From trial explorations, we expect that these algorithms should also perform well
when the discrete symmetry is more complicated than $\mathbb{Z}_2$.
The reader is also referred to \cite{Betzler:2020rfg,Bao:2020nbi,He:2020eva} for the usage of ML in recognizing dualities in QFTs.

\paragraph{}
We also consider another possible utility of ML algorithms, now in the context of the
conformal bootstrap. The state of the art numerical method involves reducing the crossing 
equations to a semi-definite programming problem which is then solved numerically
\cite{Simmons-Duffin:2015qma,Landry:2019qug}, see \cite{Poland:2018epd} for a review.
While this
yields extremely accurate results on the numerical bounds of CFT data, it can be computationally expensive. 
In this case,
can machine-learning methods assist us in extracting more about the spectrum of the CFT? We can answer this question
in the affirmative, at least for the $[\sigma\sigma]_0$ and $[\sigma\epsilon]_0$ operator
families in the 3d Ising Model, where machine-learning methods are able to train on the existing numerical data
and provide regression curves with good extrapolation. Though this is a baby example where the curve is 
somewhat simple, and indeed even the analytic expressions are known, it is an important proof-of-concept.
We especially emphasize here that machine-learning regression analysis does \textit{not} require us to input a 
form of the trial curve by hand. One may therefore expect that we could carry out a similar analysis 
even when the form of the data set is more complicated and it is not obvious if a standard curve-fitting
based regression analysis would be feasible. Indeed, these are precisely the cases where machine-learning methods 
shine, see for example \cite{Ashmore:2019wzb}, where simple, generic neural networks improve the known Donaldson algorithm for numerical Calabi-Yau metrics by almost 2 orders of magnitude.

\paragraph{}
Our motivation for these analyses is the fact that machine learning has made
important contributions in various physical and mathematical problems in
the recent past,
see \cite{MEHTA20191,He:2018jtw,Bao:2020sqg,RUEHLE20201} for reviews and references. 
In addition, machine learning methods
have now started to
routinely compete with and even outperform existing numerical and analytical methods in 
mathematics and physics \cite{sirignano2018dgm,Piscopo:2019txs,Ashmore:2019wzb,broecker2017machine}. 
It would therefore be of great
interest to apply machine learning methods to analyze CFTs. However, at
the same time one must note that often failure modes of machine learning
are not very well understood, and it is \textit{a priori} not obvious
whether a problem is machine learnable or not. The computations we present here 
indicate, at least at the level of 
proof-of-concept, that machine learning methods can be useful tools to have at hand when exploring
conformal field theories.

\paragraph{}

Finally, the results we have presented in this paper
are obtained using mainly neural networks, both for classification and regression problems. This is 
an idiosyncratic choice made mainly
for reasons of definiteness, and eventually we hope to analyse truly large amounts of CFT data using 
these methods
in near future. Neural networks are a flexible framework which scale well with large amounts of data.
More traditional machine learning methods such as support vector machines and decision trees
do also perform comparably well on the problems we have addressed here.

\paragraph{}
We provide a summary of the classification and regression problems we have carried out in Table \ref{tab:overview}.
\begin{table}
\centering
\begin{tabular}{|c|c|}
\hline
Classification Problem  & Accuracy at 80/20 Split\\ \hline\hline
Even/Odd Functions  & $100\%$\\ 
Plane/Antipodal/Null Reflection Symmetries &  $99.66\%$\\ 
Scale vs Conformal Invariance: three-point functions &  $89.99\%$\\
Scale vs Conformal Invariance: two \& three-point functions & $99.00\%$\\
Recognizing Crossing Symmetric four-point functions & $100\%$\\
Fourier Series Expansions & $98.00\%$\\
Conformal Block Expansions & $97.00\%$\\
$\mathbb{Z}_2$ symmetry of the Ising Model & $100\%$ \\
$\mathbb{Z}_3$ discrete symmetry &  $99.73\%$\\
\hline
\end{tabular}
\centering
\begin{tabular}{|c|c|}
\hline
Regression Problem & R-squared Value \\ \hline\hline
Predicing value of Fourier Coefficient & 0.9999\\ \hline
Predicting value of OPE coefficient  & 0.9988 \\ \hline
Predicting $\tau_{[\sigma\sigma]_0}(\bar{h})$ for the 3d Ising model & 
\makecell{tanh ansatz: 1.0\\ Neural Network: 0.9804} \\ \hline
Predicting $\tau_{[\sigma\epsilon]_0}(\bar{h})$ for even-spin operators &
\makecell{tanh ansatz: 1.0\\ Neural Network: 0.9966} \\ \hline
Predicting $\tau_{[\sigma\epsilon]_0}(\bar{h})$ for odd-spin operators &
\makecell{tanh ansatz: 1.0\\ Neural Network: 0.9926} \\
\hline
\end{tabular}
\caption{The classification and regression problems done in this paper evaluated by accuracy and r-squared value
respectively. A value of r-square close to one indicates a good prediction by the regressor.}
\label{tab:overview}
\end{table}

\textbf{Note:} While preparing this draft, we learned of the interesting paper \cite{Krippendorf:2020gny}
which also studies identification of symmetries in data sets by means of neural networks. Like the authors
of \cite{Krippendorf:2020gny}, we also find principal component analysis to be a useful tool to gain insight
into symmetries.
It would be interesting
to compare the two approaches in greater detail, and we hope to return to this question in the future.

\section{Detecting Symmetries using Neural Networks}\label{Sec:Detecting Symmetries}
\paragraph{}
We begin with demonstrating how neural networks may be trained to detect the presence of various discrete 
symmetries in real valued functions. To the best of our knowledge, these observations, which pave the way to our study of scale and conformal symmetries
in the correlation functions, are new.
\paragraph{}
As an aside, we would like to first make
the following observation. Typically in `visual' classification problems
that neural networks are trained for, e.g. cat vs dog classifiers, or face 
detection and recognition, the neural network has multiple hidden layers
and typically also convolutional layers (see e.g. \cite{726791,10.5555/2999134.2999257})
in order to
have the best possibility to capture fine local correlations in the image,
and to transfer learning from one patch in the image to another
patch and so on. In the classification problems we study here, there are
no local correlations that we study or allow for. The functions are 
randomly generated point clouds that obey particular global properties.
This is done with a view to ensuring that the classifier indeed learns 
the given global properties of the function we are trying to teach it, and
is not reliant on spurious or accidental local correlations. This also
leads us to expect that we could successfully work with shallow 
fully connected neural networks for
our classification tasks. Indeed, such networks prove to suffice
in what follows.

\subsection{Classifying Discrete Symmetries}\label{Subsec:Discrete}
\begin{figure}
   \begin{subfigure}[b]{0.5\linewidth}
    \includegraphics[width=\linewidth]{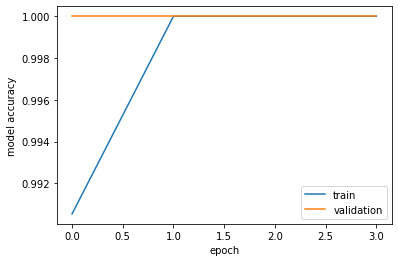}
    \caption{Accuracy}
  \end{subfigure}
  \begin{subfigure}[b]{0.5\linewidth}
    \includegraphics[width=\linewidth]{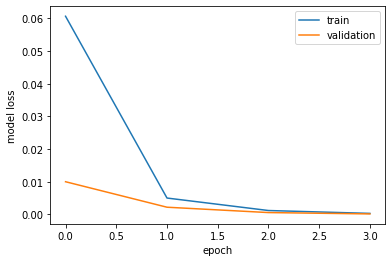}
    \caption{Loss}
  \end{subfigure}
  \caption{The Accuracy and Loss curves for the Even-Odd Classifier
  Neural Network. The data set was organized into training and validation sets in an 80/20 split.
  }
  \label{fig:even_odd}
\end{figure}

\begin{figure}
   \begin{subfigure}[b]{0.5\linewidth}
    \includegraphics[width=\linewidth]{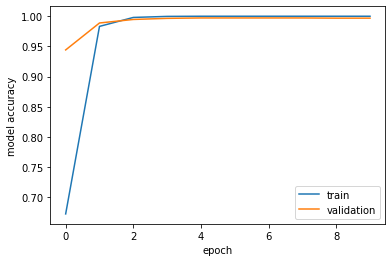}
    \caption{Accuracy}
  \end{subfigure}
  \begin{subfigure}[b]{0.5\linewidth}
    \includegraphics[width=\linewidth]{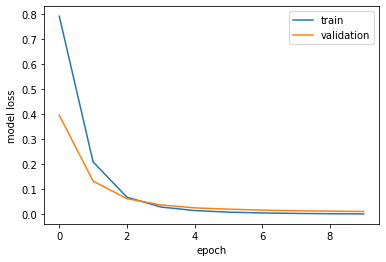}
    \caption{Loss}
  \end{subfigure}
  \caption{The Accuracy and Loss curves for the Neural Network learning the point cloud symmetries
  of Table \ref{tab:3dclassifier}.  The data set was organized into training and validation sets in an 80/20 split.}
  \label{fig:3dCloud}
\end{figure}

\begin{figure}
  \centering
  \begin{subfigure}[b]{0.4\linewidth}
    \includegraphics[width=\linewidth]{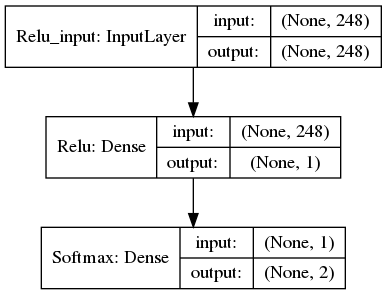}
    \caption{Even-Odd Classifier}
    \label{nn:evenodd}
  \end{subfigure}
  \begin{subfigure}[b]{0.4\linewidth}
    \includegraphics[width=\linewidth]{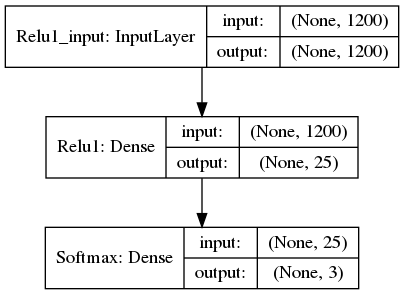}
    \caption{3d Symmetry Classifier}
    \label{nn:3dsymm}
  \end{subfigure}
  \caption{The Neural Networks for classifying even/odd functions and
  three-dimensional symmetries. The first has a total of 253 trainable
  parameters, while the second has a total of 30,103 trainable
  parameters.}
  \label{fig:NN_discrete_classifiers}
\end{figure}

\paragraph{}
One of the simplest symmetry properties a real-valued mono-variate function
may have is to be even or odd under the change of sign of its argument. We therefore consider two classes of functions, \textit{viz.}
\begin{equation}
    \left\lbrace f(x): f(-x)=f(x) \right\rbrace
    \qquad \text{and} \qquad
    \left\lbrace f(x): f(-x)=-f(x) \right\rbrace\,,
\end{equation}
and train a neural network to classify functions
under this property. The training data for the neural network is defined in Table \ref{tab:evenodddata}.
\begin{table}
\centering
\begin{tabular}{|c|c|c|}
\hline
$\mathrm{Point\,Cloud}$ & $\mathrm{X}$ & $\mathrm{Y}$ \\ \hline
$\mathrm{Even\,Function}$ &
$\left(\mathbf{x},\mathbf{f(x)},\,-\mathbf{x},\mathbf{f(x)}\right)$ & 
$\left(1,0\right)$\\ 
$\mathrm{Odd\,Function}$ &
$\left(\mathbf{x},\mathbf{f(x)},\,-\mathbf{x},-\mathbf{f(x)}\right)$ &
$\left(0,1\right)$\\ \hline
\end{tabular}
\caption{The training data for the even-odd classification
problem. $\mathbf{x}$ and $\mathbf{f(x)}$ are 64 dimensional vectors of normally distributed random numbers taking 
values in $\left[0,\pi\right]$ and $\left[-1,1\right]$ respectively. 
The neural network was trained on 10000 instances of this data.}
\label{tab:evenodddata}
\end{table}
The Neural Network trained on this data set is described in Figure
\ref{nn:evenodd}. It was trained for 4 epochs and achieved a hundred
percent accuracy. The accuracy and loss curves are shown in Figure
\ref{fig:even_odd}.
As an extension of the above classification, 
one may similarly consider classes of three-dimensional point clouds with
the various symmetry properties. 
Firstly, we consider reflection symmetry
about the $x-y$ plane, secondly, we consider antipodal reflection symmetry and
finally, point clouds which are completely randomly generated, with no
symmetries. 
The purpose of having a null case
for the classifier is to test that it is indeed capable of distinguishing the point clouds with definite symmetry properties and those without.
These symmetries and the corresponding data set are enumerated 
in Table \ref{tab:3dclassifier}, and the corresponding neural
network classifier is summarized in Figure \ref{nn:3dsymm}.
The Neural Network classifier could classify point clouds to 0.995 accuracy, as may be seen from
the accuracy curve in Figure \ref{fig:3dCloud}.
\paragraph{}
We remark, whilst on the subject of simple discrete symmetries, that we can try an even simpler problem.
Can a classifier learn modulo arithmetic?
Suppose we {\it fix} a number $p$ and establish a labelled data-set $\{n_i\} \rightarrow n \mod p $, where $n_i$ is the list of digits of $n$ in some base (it turn out that which base is not important here).
A simple classifier such as logistic regression, or a neural network such as a multi-layer perceptron with a linear layer and a sigmoid layer will very quickly ``learn'' this to accuracy and confidence very close to 1 for $p=2$ (even/odd). The higher the $p$, the more the categories to classify, and the accuracies decrease as expected.
Furthermore, if we do not fix $p$ and feed the classifier with pairs $(n,p)$ all mixed up, then, the accuracies are nearly zero. This is very much in line with the experiments of \cite{He:2017set} that the moment primes are involved, simple neural-network techniques expectantly fail in recognizing any patterns.
Nevertheless, in experiments in finite groups theory, and in arithmetic geometry have met with greater success \cite{Alessandretti:2019jbs,He:2019nzx}.

\begin{table}
\centering
\begin{tabular}{|c|c|c|}
\hline
Point Cloud Symmetry & $\mathrm{X}$ & $\mathrm{Y}$ 
\\ \hline
Plane Reflection &
$\left(\mathbf{x},\mathbf{y},\mathbf{z}\right)\,\cup\,
\left(\mathbf{x},\mathbf{y},-\mathbf{z}\right)$ & $\left(1,0,0\right)$\\ 
Antipodal Reflection &
$\left(\mathbf{x},\mathbf{y},\mathbf{z}\right)\,\cup\,
\left(-\mathbf{x},-\mathbf{y},-\mathbf{z}\right)$ & 
$\left(0,1,0\right)$\\
Null &
$\left(\mathbf{x}_1,\mathbf{y}_1,\mathbf{z}_1\right)\,\cup\,
\left(\mathbf{x}_2,\mathbf{y}_2,\mathbf{z}_2\right)$ & 
$\left(0,0,1\right)$\\\hline
\end{tabular}
\caption{The training data for the three-dimensional
point cloud classification
problem. $\mathbf{x}$, $\mathbf{y}$ and $\mathbf{z}$ are 200 dimensional vectors of normally distributed random numbers taking 
values in $\left[-1,1\right]$. The neural network was trained on 10000 instances of this data.}
\label{tab:3dclassifier}
\end{table}

\subsection{Scale Invariance vs Conformal Invariance}\label{Subsec:Scale-Conformal}
\begin{figure}
  \centering
  \begin{subfigure}[b]{0.45\linewidth}
    \includegraphics[width=\linewidth]{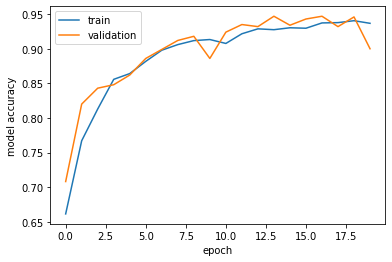}
    \caption{Accuracy}
  \end{subfigure}
  \begin{subfigure}[b]{0.45\linewidth}
    \includegraphics[width=\linewidth]{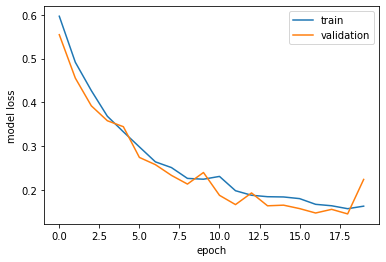}
    \caption{Loss}
  \end{subfigure}
  \caption{The Accuracy and Loss curves for the Scale vs Conformal Invariance classification when the Neural Network in Figure
  \ref{fig:NN 3pt Classifiers} is fed the conformal dimensions along
  with the corresponding three-point correlation functions. Training and validation sets are constructed with an 80/20
  split.
 }
  \label{fig:3ptCurves_1}
\end{figure}

\begin{figure}
  \centering
      \includegraphics[width=0.5\linewidth]{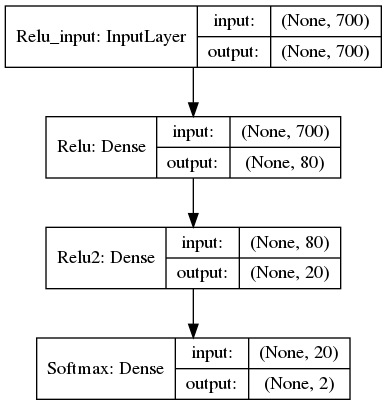}
  \caption{The Neural Network Classifier for the Scale invariant vs
  Conformally invariant three-point functions. It has 57,742 trainable parameters. 
  It processes input data of 700 dimensional feature vectors $X$, which
  are passed to a fully connected layer of 80 relu-activated neurons, whose output
  feeds to a fully connected layer of 20 relu-activated neurons. The output of this
  layer is fed to a softmax layer which outputs two possible values for $Y$. `None' is a 
  placeholder to indicate that the neural network trains on an arbitrary number of training
  instances $(X,Y)$.}
  \label{fig:NN 3pt Classifiers}
\end{figure}

\paragraph{}
We next turn to study how neural networks may be trained to distinguish between scale invariance 
and conformal invariance from the correlation functions exhibiting either of these symmetries.
This can be useful for machine to distinguish the emergence of scale or conformal symmetries in condensed matter systems.
\paragraph{}
Let us briefly review the relevant details about scale and conformal invariant correlation functions we will perform the machine-learning. 
Given a quantum field theory
in $d$-dimensional space-time, containing scalar operators $\co_1$, $\co_2$ and $\co_3$ of scaling dimensions 
$\Delta_1$, $\Delta_2$ and $\Delta_3$ respectively, the two point function is constrained by 
scale invariance to be of the form \cite{Nakayama:2013is}:
\begin{equation}
\left\langle\co_{\Delta_i}\left(x_1\right)\co_{\Delta_j}\left(x_2\right)\right\rangle
=\frac{c_{ij}}{\left\vert x_{12}\right\vert^{\Delta_i+\Delta_j}}\,, \quad x_{ij}=x_i-x_j,
\end{equation}
where $|x_{ij}| = \sqrt{(x_i-x_j)^2}$ and $c_{ij}$ is an overall normalization constant.
Requiring conformal invariance imposes the additional two point orthogonality condition such that 
$c_{ij}=c^{(i)}\delta_{ij}$. 
Additionally, conformal invariance imposes even more restrictive constraints on the three-point correlation functions of the theory.
In particular, if the theory has only scale invariance, the form of the three-point function is fixed to be:
\begin{equation}\label{eq:3ptscale}
\begin{split}
&\left\langle\co_{\Delta_1}\left(x_1\right)\co_{\Delta_2}\left(x_2\right)
\co_{\Delta_3}\left(x_3\right)\right\rangle=
 \sum_{\{a,b,c\}} {c_{abc}\over\left|x_{12}\right|^{a}\left|x_{23}\right|^{b}
\left|x_{31}\right|^{c}}\,,\\& 
\qquad a+b+c= \Delta_1+\Delta_2+\Delta_3\,,
\end{split}
\end{equation}
where $a$, $b$ and $c$ are positive real numbers, constrained as above.
On the other hand, if the theory has full conformal invariance, then 
exponents of $x_{ij}$ in the three-point function are completely
determined, i.e.
\begin{equation}\label{eq:3ptCFT}
\left\langle\co_{\Delta_1}\left(x_1\right)\co_{\Delta_2}\left(x_2\right)
\co_{\Delta_3}\left(x_3\right)\right\rangle
= {\lambda_{123}\over\left|x_{12}\right|^{\alpha_{123}}\left|x_{23}\right|^{\alpha_{231}}
\left|x_{31}\right|^{\alpha_{312}}}\,,
\end{equation}
where $\alpha_{ijk}=\Delta_i+\Delta_j-\Delta_k$ and $\lambda_{ijk}$ is known as the OPE coefficient.
\paragraph{}
We shall begin with a simple classification problem, where we 
supply the neural network outlined in Figure 
\ref{fig:NN 3pt Classifiers} a point cloud of 
positions $x_1$, $x_2$, $x_3$, and
quantum numbers $\Delta_i$
of the three operators. We consider the two classes of functions
given in Equations \eqref{eq:3ptscale} and \eqref{eq:3ptCFT} and 
train the classifier to distinguish between the two functions. 
Our training set is explained in greater detail in 
Table \ref{tab:3ptclassifier1}, and the training curves for the neural
network are shown in Figure \ref{fig:3ptCurves_1}, from where it is
apparent that the classifier can easily train up to around 95 percent 
accuracy.
Subsequently, we generated a new set of testing data of three-point functions
for the classifier, and found that the neural network could classify these
functions, which had not been used for training or validation, 
to 97.15
percent accuracy.
\begin{table}
\centering
\begin{tabular}{|l|c|r|}
\hline
$\mathrm{Symmetry}$ & $\mathrm{X}$ & $\mathrm{Y}$ 
\\ \hline
$\mathrm{Scale\, Invariance}$ &
$\left(\mathbf{x}_{12},\mathbf{x}_{23},\mathbf{x}_{31},\Delta_1,
\Delta_2,\Delta_3,f^{(3)}_{scale}\right)_{(100)}$ & $\left(1,0\right)$\\ 
$\mathrm{Conformal\, Invariance}$ &
$\left(\mathbf{x}_{12},\mathbf{x}_{23},\mathbf{x}_{31},\Delta_1,
\Delta_2,\Delta_3,f^{(3)}_{cft}\right)_{(100)}$ & 
$\left(0,1\right)$\\
\hline
\end{tabular}
\caption{The first set of training data for the classification
of scale and conformal invariant correlation functions. 
$x_{ij}$ is the invariant distance $\sqrt{\left(x_i-x_j\right)^2}$
and the $\Delta_i$ are the conformal dimensions of the operators
$\mathcal{O}_i$ positioned at $x_i$. The three-point
functions $f^{(3)}_{cft}$
and $f^{(3)}_{cft}$ are defined in \eqref{eq:3ptscale} and 
\eqref{eq:3ptCFT} respectively. For each training instance,
$a$, $b$ and $c$ are chosen randomly to satisfy the constraint
in \eqref{eq:3ptscale}. The subscript (100) indicates that
each such instance of training data is a $100\times 7$ array.
This is flattened into a 700 dimensional vector and fed to the neural
network along with its value of $Y$ as a single training instance.
The neural network was trained on 10000 instances of this data, i.e.
5000 scale invariant functions and 5000 conformal invariant functions.}
\label{tab:3ptclassifier1}
\end{table}
\paragraph{}
It turns out that one may improve the classifier performance by an
alteration in how the training data is presented to it.
Next, rather than supplying the classifier the conformal dimensions of operators
and the three-point function,
we supplied it two-point functions 
of each operator, along with the three-point function.
The neural network could distinguish between a randomly generated set of
scale and conformally invariant correlation functions that it had not been
trained on to 99.00 percent accuracy. 
\begin{table}
\centering
\begin{tabular}{|c|c|c|}
\hline
$\mathrm{Symmetry}$ & $\mathrm{X}$ & $\mathrm{Y}$ 
\\ \hline
$\mathrm{Scale\, Invariance}$ &
$\left(\mathbf{x}_{12},\mathbf{x}_{23},\mathbf{x}_{31},
f^{(2)}_{\Delta_1}, f^{(2)}_{\Delta_2},f^{(2)}_{\Delta_3},
f^{(3)}_{\rm scale}\right)_{(100)}$ & $\left(1,0\right)$\\ 
$\mathrm{Conformal\, Invariance}$ &
$\left(\mathbf{x}_{12},\mathbf{x}_{23},\mathbf{x}_{31},
f^{(2)}_{\Delta_1}, f^{(2)}_{\Delta_2},f^{(2)}_{\Delta_3},
f^{(3)}_{\rm conformal}\right)_{(100)}$ & 
$\left(0,1\right)$\\
\hline
\end{tabular}
\caption{The second set of training data for the classification
of scale and conformal invariant correlation functions. Instead of
passing the conformal dimensions $\Delta_i$ as done previously,
we pass it the point cloud of the respective two point correlation
functions.
The neural network was trained on 10000 instances of this data, i.e.
5000 scale invariant functions and 5000 conformal invariant functions.}
\label{tab:3ptclassifier2}
\end{table}
\begin{figure}
  \centering
  \begin{subfigure}[b]{0.45\linewidth}
    \includegraphics[width=\linewidth]{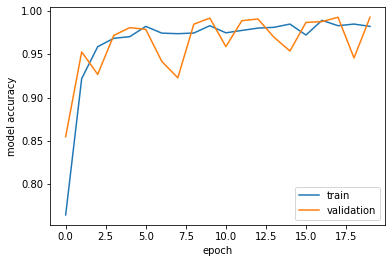}
    \caption{Accuracy}
  \end{subfigure}
  \begin{subfigure}[b]{0.45\linewidth}
    \includegraphics[width=\linewidth]{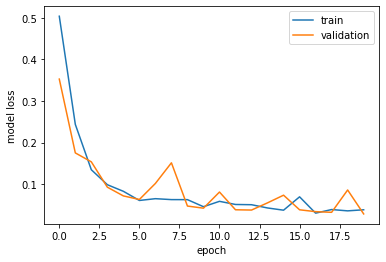}
    \caption{Loss}
  \end{subfigure}
  \caption{The Accuracy and Loss curves for the Scale vs Conformal Invariance classification when the Neural Network in Figure
  \ref{fig:NN 3pt Classifiers} is fed a point-cloud of two-point 
  and three-point correlation functions. Training and validation sets are constructed with an 80/20 split.}
  \label{fig:3ptCurves_2}
\end{figure}
The learning curves are in Figure \ref{fig:3ptCurves_2}.
The takeaway from these computations seems to be that neural network
classifiers can quite easily distinguish scale invariance from conformal
invariance in the theory from the knowledge of the conformal dimensions
or two point functions, as well as three point functions.
It would be interesting to see if a similar classification is successful
for spinning three-point correlation functions.

\subsection{Crossing Symmetric Four-Point Functions}\label{Subsec:Crossing-4pt}
\begin{figure}
  \centering
      \includegraphics[width=0.5\linewidth]{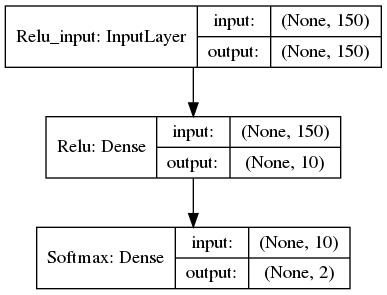}
  \caption{The Neural Network Classifier for recognizing crossing symmetry. This
  has 1532 trainable parameters.}
  \label{fig:NN 4pt Classifiers}
\end{figure}

\begin{figure}
  \centering
  \begin{subfigure}[b]{0.45\linewidth}
    \includegraphics[width=\linewidth]{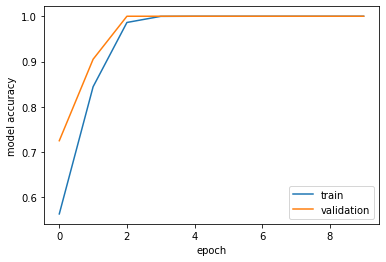}
    \caption{Accuracy}
  \end{subfigure}
  \begin{subfigure}[b]{0.45\linewidth}
    \includegraphics[width=\linewidth]{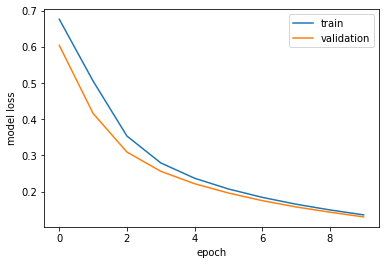}
    \caption{Loss}
  \end{subfigure}
  \caption{The Accuracy and Loss curves for the crossing symmetry
  classification when the Neural Network in Figure
  \ref{fig:NN 4pt Classifiers} is fed the data in Table
  \ref{tab:4ptclassifier} with an 80/20 split for training and validation.
  }
  \label{fig:4ptCurves}
\end{figure}

\paragraph{}
Given a four-point correlation function of identical scalar primary operators in a CFT
\begin{equation}
    \left\langle \mathcal{O}_1\left(x_1\right)
    \mathcal{O}_2\left(x_2\right) \mathcal{O}_3\left(x_4\right)
    \mathcal{O}_4\left(x_4\right)\right\rangle
    \equiv f\left(x_1,x_2,x_3,x_4\right)\,,
\end{equation}
where we have again suppressed the dependence on conformal dimensions of the 
external operators,
crossing symmetry is the property that:
\begin{equation}
    f\left(x_1,x_2,x_3,x_4\right)=f\left(x_1,x_3,x_2,x_4\right)
    =f\left(x_1,x_4,x_3,x_2\right)\,.
\end{equation}
In the following we shall train a neural network classifier to recognize
a four point function which is crossing symmetric, as opposed to a generic
function of four variables which has no definite crossing symmetry 
property. Our overall strategy is the same as before. 
We will feed a data set
of point clouds with and without this symmetry and train the neural
network to distinguish between them. In practice, however, we achieve
much better results by training the neural network to recognize the
following expression
\begin{equation}
\begin{split}
    f\left(x_1,x_2,x_3,x_4\right)
    &=f\left(x_1+\epsilon_1,x_3+\epsilon_3,x_2+\epsilon_2,x_4+\epsilon_4\right)
    +\mathcal{O}\left(\epsilon\right)\\
    &=f\left(x_1-\epsilon_1,x_4-\epsilon_4,x_3-\epsilon_3,x_2-\epsilon_2\right)
    +\mathcal{O}\left(\epsilon\right)\,.
\end{split}
\end{equation}
Here $\epsilon_i$ may be viewed heuristically as a cutoff, whose role is
not fully clear \textit{a priori}. Its appearance indicates that for a neural network 
to recognize crossing symmetry, we need to study the crossing equation
not just at the interchanged arguments, but in their neighbourhoods,
the underlying reason for this is currently not obvious to us.
In practice, we achieved good results by sampling point
clouds of 100 points to define a training instance, and keeping 
$\epsilon\simeq 0.05$.
The data set is enumerated in Table \ref{tab:4ptclassifier}.
\begin{table}
\centering
\begin{tabular}{|l|c|r|}
\hline
$\mathrm{Symmetry}$ & $\mathrm{X}$ & $\mathrm{Y}$ 
\\ \hline
$\mathrm{Crossing\,\, Symmetric}$ &
\makecell{$\big(\mathbf{x}_{1234},
f^{(1234)},\mathbf{x}_{1324}+\epsilon,\,f^{(1234)}+\mathcal{O}(\epsilon)$,
\\ $\mathbf{x}_{1432}+\epsilon,\,f^{(1234)}+\mathcal{O}(\epsilon)\big)$}
& $\left(1,0\right)$\\ 
$\mathrm{Non-Symmetric}$ &
\makecell{$\big(\mathbf{x}_{1234},
f^{(1234)},\mathbf{x}_{1324}+\epsilon,\,f^{(1324)}+\epsilon$,\\ 
$\mathbf{x}_{1432}+\epsilon,\,f^{(1423)}+\epsilon\big)$}& 
$\left(0,1\right)$\\
\hline
\end{tabular}
\caption{The training data for the recognition of crossing symmetric
four-point functions. The neural network was trained on 1000 instances of this data, i.e.
500 crossing symmetric functions and 500 non-symmetric functions. $x_{ijkl}$ denotes the quadruplet
$(x_1,x_j,x_k,x_l)$ and $f^{(ijkl)}$ denotes the value of the function $f$ sampled at the arguments $x_{ijkl}$.}
\label{tab:4ptclassifier}
\end{table}
The architecture of this classifier is sketched in Figure 
\ref{fig:NN 4pt Classifiers}
and the learning curves are available in Figure \ref{fig:4ptCurves}.
We have now seen that two important characteristics of CFT correlation functions, namely the conformal invariance of the 
three-point function and the crossing symmetry of the four-point function, can be recognized to great accuracy
by neural networks. We shall next turn our attention to training the machine to recognize another characteristic of CFT four-point functions, i. e. their operator product expansion decomposition into individual conformal blocks.

\section{Machine Learning The Conformal Block Expansion}\label{Section:Conformal Expansion}
\paragraph{}
In this section we shall consider some methods by which one may try to machine learn the conformal
block expansion of a four-point function in a CFT. For definiteness and simplicity, we shall stick to
an one-dimensional CFT, where the four-point function of identical scalar primary operator $\phi_{h_\phi}(x)$ may be decomposed as:
\begin{equation}\label{eq:4ptCFT1}
    \langle\phi_{h_\phi}(x_1) \phi_{h_\phi}(x_2) \phi_{h_\phi}(x_3) \phi_{h_\phi}(x_4) \rangle = \frac{F_{\phi\phi\phi\phi}(z)}{|x_{12}|^{2h_{\phi}}|x_{34}|^{2h_\phi}}= \frac{\sum_{\lbrace h\rbrace} c_h\,z^h\,_{2}F_{1}\left(h,h,2h;z\right)}{|x_{12}|^{2h_{\phi}}|x_{34}|^{2h_\phi}}\,.
\end{equation}
\begin{equation}
z = \frac{|x_{12}||x_{34}|}{|x_{13}||x_{24}|}.   
\end{equation}
The function $z^h\,_{2}F_{1}\left(h,h,2h;z\right)$ is the one-dimensional conformal block, which only depends on a single conformal cross ratio $z$ and is labeled by scaling dimension $h$ of the exchange operator $\mathcal{O}_h$, and $c_h$ is the associated OPE coefficient. 
\paragraph{}
Given the expansion \eqref{eq:4ptCFT1}, two interesting questions naturally suggest themselves. Firstly, can neural networks be
trained to detect the
presence (or conversely the absence) of a conformal block belonging to a particular primary operator in the above expansion; and secondly,  can machine be trained
to estimate the value of the OPE coefficients $\{c_h\}$ for a given $F_{\phi\phi\phi\phi}(z)$. We shall provide supporting evidence for the affirmative answers to both questions.
It is interesting to note that in the so-called ``diagonal limit'' where the conformal cross ratios $z =\bar{z}$ which is of interests for conformal bootstrap computations, the general d-dimensional conformal block can be expressed in a closed form in terms of a finite sum of generalized hypergeometric function ${}_3 F_2$ \cite{Hogervorst:2013kva}, we thus expect our analysis here can be readily adapted for more general situations \footnote{Moreover, the general $z \neq \bar{z}$ d-dimensional conformal blocks can be expressed in terms of a finite sum of Appell's hypergeometric function $F_4$ \cite{Chen:2019gka}.}. 
However we should also make a disclaimer here that we have merely regarded the conformal block as a complete kinematic basis, without imposing the additional consistency condition such as crossing symmetry, so this should only be taken as the first step towards using machine-learning method to perform conformal bootstrap.

\subsection{A Warm-Up with the Fourier Expansion}\label{Subsection:Fourier Expansion}
\begin{figure}
  \centering
  \begin{subfigure}[b]{0.45\linewidth}
    \includegraphics[width=\linewidth]{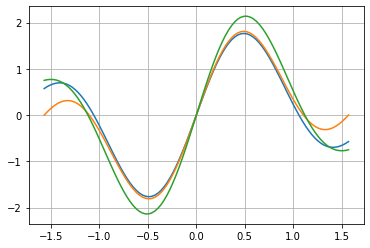}
    \caption{Functions with $a_n \neq 0$.}
  \end{subfigure}
  \begin{subfigure}[b]{0.45\linewidth}
    \includegraphics[width=\linewidth]{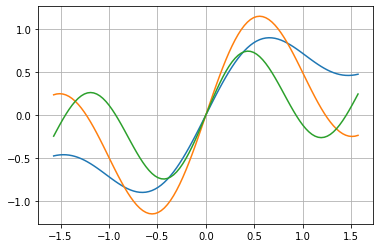}
    \caption{Functions with $a_2 = 0$.}
  \end{subfigure}
  \caption{Functions expandable in terms of the first four sine modes. Different colors correspond to different
  functions.}
  \label{fig:fourier classification}
\end{figure}

\begin{figure}
  \centering
  \begin{minipage}[t]{.55\linewidth}
    \subcaptionbox{The neural network architecture for the classifier. \label{fig:FourierClassifierNN}}
      {\includegraphics[width=\linewidth]{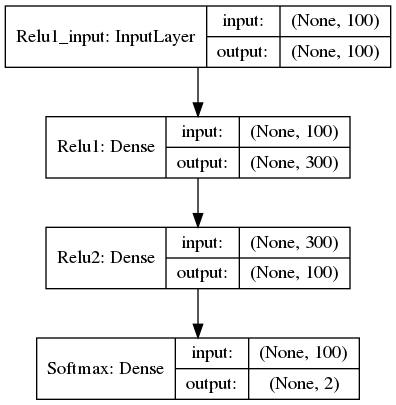}}%
  \end{minipage}%
  \hfill
  \begin{minipage}[b]{.45\linewidth}
    \subcaptionbox{\small{Accuracy}  \label{fig:fourierclassificationaccuracy}}
      {\includegraphics[width=\linewidth]{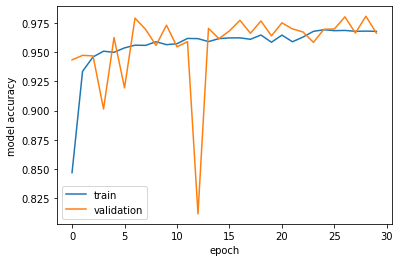}}
      
    \subcaptionbox{Loss \label{fig:fourierclassificationloss}}
      {\includegraphics[width=\linewidth]{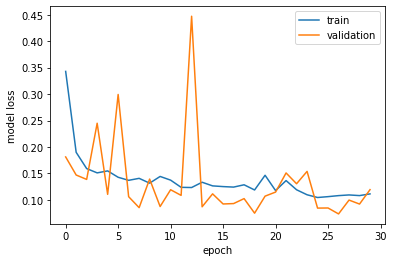}}%
  \end{minipage}%
  \caption
    {%
      Neural Network Classification of functions by their Fourier sine expansions. The 100 dimensional 
      feature vector
      defined in Table \ref{tab:fourierclassifier} is fed to a two layers consisting 
      of 300 and 100 relu-activated neurons respectively. The output is from the subsequent softmax layer.
    }%
\end{figure}

\begin{figure}
  \centering
      \includegraphics[width=0.6\linewidth]{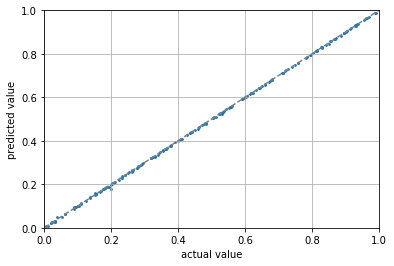}
  \caption{A comparison of the actual values of Fourier sine mode $a_2$ for 200 test functions with the corresponding
  predicted values. For a regressor working at 100 percent accuracy, the scatter points would lie precisely on the
  diagonal line $y=x$, shown as a dashed line here. We obtained an r-squared value of 0.9999.}
  \label{fig:FourierRegressorValues}
\end{figure}

\paragraph{}
We begin
with an \textit{a priori} simpler case of the Fourier decomposition in one dimension
as a warm-up to the conformal block decomposition and demonstrate how various aspects of the 
Fourier expansion are 
machine-learnable. In particular, consider a function with a domain
$x\in\left(-\frac{\pi}{2},\frac{\pi}{2}\right)$ that admits a purely sine expansion
\begin{equation}\label{eq:fourier}
    f\left(x\right) = \sum_{n=1}^\infty a_n\,\sin\left(n\,x\right)\,.
\end{equation}
We will train a neural network to answer the following two questions. Firstly, to identify
a class of functions for which an arbitrary but fixed Fourier mode is missing in the expansion
\eqref{eq:fourier}. Subsequently, we shall turn our attention to 
a harder regression problem, which is to have the neural network output the value of a fixed sine
coefficient, given values of the function sampled in the interval $\left(-\frac{\pi}{2},\frac{\pi}{2}\right)$.
\paragraph{}
We should emphasize that the main goal of this computation is not to obtain a neural network implementation
of the Fourier sine transform. 
Instead, we argue that by doing these classification and regression tasks accurately, we demonstrate
neural networks are essentially able to identify fine-grained information about how symmetries manifest themselves
in physical quantities. Both classes of functions that appear in Equation \eqref{eq:fourier classes} carry the
same $U(1)$ symmetry, and indeed the instances plotted in Figure \ref{fig:fourier classification} look roughly
similar at first glance. But, to a neural network
classifier, these functions are vastly different. This is further borne
out by the Principal Component analysis that we will later present. The same sensitivity to the representations
of symmetry that appear in the expansion of a function is also seen later in the context of the conformal block 
expansion. Highlighting this sensitivity is the main goal of this section.
\paragraph{}
For definiteness, we shall pick the $n=2$ mode for both the classification, and the regression tasks. 
Also, while a generic function $f(x)$ has an infinite number of sine modes, we shall consider a sub-class
of functions for which $a_{n\geq 5}=0$. This is both because considering this sub-class is sufficient to illustrate
the concepts that follow, and also because
as we include higher and higher sine modes, the 
structure of the neural network required to accurately carry out these tasks 
becomes more and more complicated.
\paragraph{}
Beginning with the classification problem, we have two classes of functions, namely
\begin{equation}\label{eq:fourier classes}
    f^{(1)}\left(x\right) = \sum_{n=1}^4 a_n\,\sin\left(n\,x\right)\,,
    \qquad
    f^{(2)}\left(x\right) = \sum_{n=1,\, n\neq 2}^4 a_n\,\sin\left(n\,x\right)\,.
\end{equation}
We sample the function at 100 uniformly spaced points from $\left(-\frac{\pi}{2},\frac{\pi}{2}\right)$.
The values of $ f^{(1)}$ and $ f^{(2)}$ at these points constitute the training data, as shown in Table
\ref{tab:fourierclassifier}.
\begin{table}
\centering
\begin{tabular}{|c|c|c|}
\hline
$\mathrm{Modes}$ & $\mathrm{X}$ & $\mathrm{Y}$ 
\\ \hline
$a_{n} \in (0,1)\,\forall\, n\leq 4$ &
$\left\lbrace f^{(1)}\left(x_i\right) \quad \vert \quad x_i = -\frac{\pi}{2}+i\,\frac{\pi}{N}\,,
i \in [0,N-1]
\right\rbrace$ & $\left(1,0\right)$\\ 
$a_2=0$ &
$\left\lbrace f^{(2)}\left(x_i\right) \quad \vert \quad x_i = -\frac{\pi}{2}+i\,\frac{\pi}{N}\,,
i \in [0,N-1]
\right\rbrace$ & 
$\left(0,1\right)$\\
\hline
\end{tabular}
\caption{The training data for the classification
of functions according to their Fourier sine modes. We took $N=100$, and
$f^{(1)}$ and $f^{(2)}$ are as defined in \eqref{eq:fourier classes}.
The neural network was trained on 200000 instances of this data, i.e.
100000 functions with all modes $a_{n\leq 4}$ randomly sampled from
a normal distribution valued in $(0,1)$ and 100000 functions
with the mode $a_{2}=0$ and the remaining modes sampled as before.}
\label{tab:fourierclassifier}
\end{table}
For illustration, we plot some functions of each class in Figure \ref{fig:fourier classification}.
From a visual inspection of the graphs, it does not seem very obvious that these functions
fall into two sharply delineated categories, but as we see from the training curves
in Figure \ref{fig:fourierclassificationaccuracy} and \ref{fig:fourierclassificationloss}, 
the neural network defined in Figure \ref{fig:FourierClassifierNN}
is able to distinguish between the two classes to 97 percent accuracy by training over 30 epochs, which takes a
few minutes on a laptop. We also observed that performance erodes rapidly as higher modes are turned on.
We expect that this is a consequence of the Universal Approximation Theorem 
\cite{Cybenko1989,HORNIK1991251}, see also Chapter 4 of \cite{nielsenneural} for a visual demonstration
of the theorem. This essentially states that while it is certainly
possible for a neural network to mimic a function with arbitrarily rich features, 
determining the necessary parameters of the network becomes an increasingly difficult task in practice. 
\paragraph{}
We subsequently tested the classifier on 2000 new functions, 1000 of each class. The classifier distinguished
them to 97 percent accuracy and the confusion matrix was
$\big(\begin{smallmatrix} 1000 & 0 \\ 70 & 930 \end{smallmatrix}\big)$, i.e. all 1000 instances of functions 
with vanishing $a_2$ were classified correctly, while 70 of the functions with all non-vanishing $a_n$ were 
mis-classified by the network, and 930 were correctly classified.
\footnote{The confusion matrix is a standard representation of how accurately a classification algorithm
works on a data set. It is defined as the matrix whose $ij$th element is the number of elements of class $j$
classified as being elements of class $i$. In classification problems with more than
two classes, it can often give insights into errors being made by the machine. See \cite{59192} for a textbook example
based on the MNIST data set \cite{lecun-mnisthandwrittendigit-2010} of handwritten digits from 0 to 9.
It should be apparent that off-diagonal elements of this matrix denote
incorrect classifications on part of the classifier, 
and a purely diagonal confusion matrix indicates that the all the
elements of the data set were classified correctly by the classifier.}
\paragraph{}
We now turn to the regression problem, i.e. given a function $f(x)$ with the sine expansion
\begin{equation}\label{eq:fourier expansion 2}
    f\left(x\right) = \sum_{n=1}^4 a_n\,\sin\left(n\,x\right)\,,
\end{equation}
we would like the neural network to output for us the coefficient $a_2$. The data set for the regression problem
is outlined in Table \ref{tab:fourier regressor}. 
\begin{table}
\centering
\begin{tabular}{|c|c|c|}
\hline
$\mathrm{Modes}$ & $\mathrm{X}$ & $\mathrm{Y}$ 
\\ \hline
$a_{n} \in (0,1)\,\forall\, n\leq 4$ &
$\left\lbrace f\left(x_i\right) 
\right\rbrace$ & 
$\left\lbrace a_2\sin \left(2\,x_i\right) \quad \vert \quad x_i = -\frac{\pi}{2}+i\,\frac{\pi}{4}\,,
i \in [0,3]
\right\rbrace$\\ 
\hline
\end{tabular}
\caption{The training data for learning the coefficient of the Fourier mode $a_2$
when supplied the function $f(x)$ defined in Equation \eqref{eq:fourier expansion 2}.
We choose $N=100$ and the $\left\lbrace f(x_i)\right\rbrace$ are sampled at the $x_i$ defined
previously in Table 
\ref{tab:fourierclassifier}.
The neural network was trained on 100000 instances of this data, i.e.
100000 functions with all modes $a_{n\leq 4}$ randomly sampled from
a normal distribution valued in $(0,1)$.}
\label{tab:fourier regressor}
\end{table}
Note that for the output $Y$, we have chosen to sample the mode $a_2\,\sin\left(2\,x_i\right)$
at four values of $x$, 
rather than directly output the value $a_2$. This is because the neural network is found to overfit
to the training set
when supplied with just the single value $a_2$. The architecture of the neural network is identical to
the classifier in Figure \ref{fig:FourierClassifierNN}, except that the softmax layer is replaced with a
layer of four output neurons. Once the regressor is trained, we test it on new functions. 
Supplied
with the four values of $a_2\,\sin\left(2\,x_i\right)$ as output from the regressor for each input $f(x)$, 
we divide
them out by $\sin\left(2\,x_i\right)$ and take the median. This is our predicted value of $a_2$. 
We compare the 
predicted value with the actual value for 200 functions, and the result is given in Figure
\ref{fig:FourierRegressorValues}. We find that the regressor is able to predict the $a_2$ value to the accuracy within
0.5 percent. Though we do not show it here, similar results hold for the remaining modes.

\subsection{The Conformal Block Expansion}\label{subsection:cb}
\begin{figure}
  \centering
  \begin{subfigure}[b]{0.45\linewidth}
    \includegraphics[width=\linewidth]{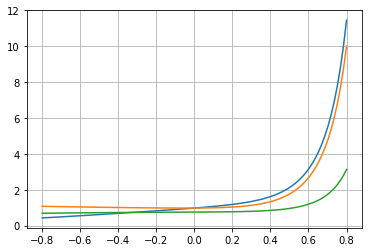}
    \caption{Functions with $c_n \neq 0$.}
  \end{subfigure}
  \begin{subfigure}[b]{0.45\linewidth}
    \includegraphics[width=\linewidth]{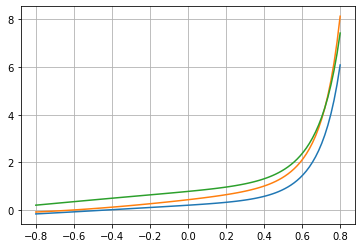}
    \caption{Functions with $c_2 = 0$.}
  \end{subfigure}
  \caption{Instances of the two classes of functions enumerated in Equation \eqref{eq:4ptCFT1 b}. Different functions
  are represented by different colors.}
  \label{fig:block classification}
\end{figure}

\begin{figure}
  \centering
      \includegraphics[width=0.5\linewidth]{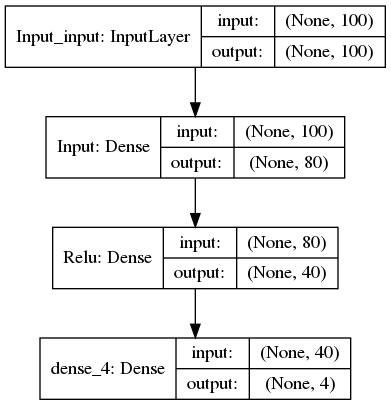}
  \caption{The neural network architecture used to perform classification and regression tasks for the conformal
  block expansion. The figure above corresponds to the regressor, which outputs the coefficient $c_2$ by giving
  values the quantity $c_2\,z^2\,{}_2F_1\left(2,2,4,z\right)$ at four different points $z$ at the output
  layer. The architecture of the classifier
  is identical, except that the output layer is now a softmax layer.}
  \label{fig:CBRegressorNN}
\end{figure}

\begin{figure}
  \centering
  \begin{subfigure}[b]{0.45\linewidth}
    \includegraphics[width=\linewidth]{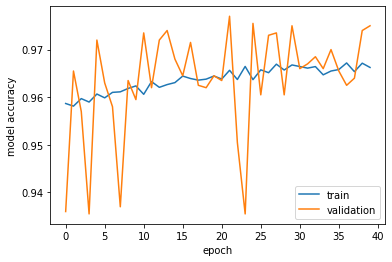}
    \caption{Accuracy}
  \end{subfigure}
  \begin{subfigure}[b]{0.45\linewidth}
    \includegraphics[width=\linewidth]{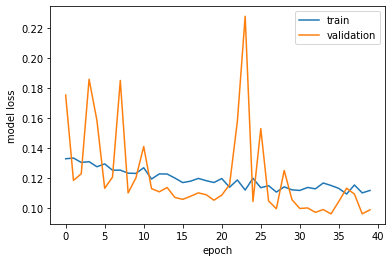}
    \caption{Loss}
  \end{subfigure}
  \caption{Accuracy and Loss curves of the Neural Network classifier for the Conformal Block expansion. }
  \label{fig:block classification training}
\end{figure}

\begin{figure}
  \centering
      \includegraphics[width=0.6\linewidth]{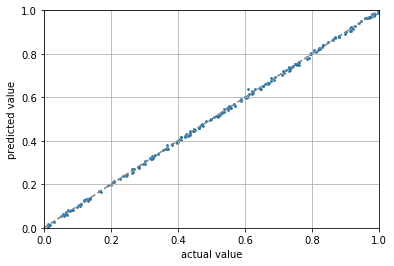}
  \caption{A comparison of the actual values of the OPE
  coefficient $c_2$ for 200 test functions with the corresponding
  predicted values. A regressor functioning at 100 percent accuracy would produce the dashed diagonal line
  corresponding to $y=x$ in the above graph. Actual results appear above as a scatter plot about that line.
  The r-squared value is 0.9988.}
  \label{fig:CBRegressorValues}
\end{figure}

\paragraph{}
With the experience of the Fourier sine expansion, let us now turn to the case of the one-dimensional conformal block
expansion. Here we will make a simplifying assumption 
that the scaling dimensions $h$ of the operators
being exchanged in the four point function \eqref{eq:4ptCFT1} are integer valued
\footnote{We have also trained the classifier to achieve the same performance
when this condition is relaxed to allow the $h_i$ to be non-integers.
However, the classifier now needs to be trained over values of 
the function \eqref{eq:4ptCFT1 b} sampled
at complex $z$ as well. For simplicity, we present the case of integer $h_i$s here.}.
This assumption may seem somewhat non-physical, however this is sufficient for demonstrating that the presence or absence of a particular conformal block in the OPE expansion is machine learnable.
Thus we start with the conformal block expansion:
\begin{equation}\label{eq:4ptCFT1 b}
    \hat{F}_{\phi\phi\phi\phi}\left(z\right) = \sum_{n=0}^{\infty} c_n\,z^n\,_{2}F_{1}\left(n,n,2n;z\right)\,.
\end{equation}
Furthermore as in the Fourier case, in practice, we truncate the sum over $n$ at a finite value $N$, which was 10 for classification
and 5 for regression. Again, we pick the $n=2$ term for both classification, and regression. Curiously, the neural
networks required to perform these tasks are much smaller than their counterparts for the Fourier case.
\begin{table}
\centering
\begin{tabular}{|c|c|c|}
\hline
OPE Coefficients & $\mathrm{X}$ & $\mathrm{Y}$ 
\\ \hline
$c_{n} \in (0,1)\,\forall\, n\leq 10$ &
$\left\lbrace f^{(1)}\left(x_i\right) \quad \vert \quad x_i = -0.8+i\,\frac{1.6}{N}\,,
i \in [0,N-1]
\right\rbrace$ & $\left(1,0\right)$\\ 
$c_2=0$ &
$\left\lbrace f^{(2)}\left(x_i\right) \quad \vert \quad x_i = -0.8+i\,\frac{1.6}{N}\,,
i \in [0,N-1]
\right\rbrace$ & 
$\left(0,1\right)$\\
\hline
\end{tabular}
\caption{The training data for the classification
of functions \eqref{eq:4ptCFT1} according to their non-vanishing OPE coefficients. 
We took $N=100$, and sampled the functions 
$f^{(1)}$ and $f^{(2)}$ defined in \eqref{eq:4ptCFT1 classification}.
The neural network was trained on 200000 instances of this data, i.e.
100000 functions with all modes $c_{n\leq 10}$ randomly sampled from
a normal distribution valued in $(0,1)$ and 100000 functions
with the mode $c_{2}=0$ and the remaining modes sampled as before.}
\label{tab:cbclassifier}
\end{table}
We begin with the classification problem, where we have two classes of functions, i.e.
\begin{equation}\label{eq:4ptCFT1 classification}
    f^{(1)}\left(z\right) = \sum_{n=0}^{10} c_n\,z^n\,_{2}F_{1}\left(n,n,2n;z\right)\,,\qquad
    f^{(2)}\left(z\right) = \sum_{n=0,n\neq 2}^{10} c_n\,z^n\,_{2}F_{1}\left(n,n,2n;z\right)\,.
\end{equation}
To build the training set, these functions were generated by randomly generating $c_n$s, normally distributed
between 0 and 1. The functions were then 
sampled at 100 uniformly spaced points in the interval $(-0.8,0.8)$, again chosen for definiteness
as the expressions are diverge at $\vert z\vert =1$, due to the presence of the hypergeometric function ${}_2F_1$
\footnote{Physically for CFT four-point functions
$z$ is restricted to lie in $0\le z\le 1$, but sampling the functions on the real line
in this range
leads to an indifferent performance of the neural networks. It is possible to improve the performance by
going to the complex plane, and indeed this is necessary if we work at non-integer weights.}.
Example functions of both classes are plotted in 
Figure \ref{fig:block classification}, and the complete training data is described in Table \ref{tab:cbclassifier}.
\paragraph{}
As in the Fourier expansion case, there does not appear to be an obvious characteristic which separates the two classes.
However, a neural network of a single layer of 40 relu activated neurons is able to distinguish the two classes to
96 percent accuracy. 
The training curves are shown in Figure \ref{fig:block classification training}
and the neural network architecture is shown in
Figure \ref{fig:CBRegressorNN}. 
We subsequently tested this trained classifier on 2000 new functions, which it was able to distinguish
again to 97 percent accuracy. 
The confusion matrix in this case was 
$\big(\begin{smallmatrix} 1000 & 0 \\ 67 & 933 \end{smallmatrix}\big)$, i.e. all 1000 cases of the functions 
$f^{(2)}(z)$
were classified correctly, while 67 of the functions $f^{(1)}(z)$ were mis-classified as $f^{(2)}(z)$.
\paragraph{}
We next turn to the regression problem for conformal blocks, which we pose as follows. Given a set of functions of the
form
\begin{equation}\label{eq:4ptCFT1 regression}
    f\left(z\right) = \sum_{n=0}^5 c_n\,z^n\,_{2}F_{1}\left(n,n,2n;z\right)\,,
\end{equation}
we would like to train a neural network to predict the value of $c_2$. Our training data is described in 
Table \ref{tab:cb regressor}.
\begin{table}
\centering
\begin{tabular}{|c|c|c|}
\hline
OPE Coefficients & $\mathrm{X}$ & $\mathrm{Y}$ 
\\ \hline
$c_{n} \in (0,1)\,\forall\, n\leq 5$ &
$\left\lbrace f\left(x_i\right) 
\right\rbrace$ & 
$\left\lbrace c_2\,z^2\,_{2}F_{1}\left(2,2,4;z\right) \quad \vert \quad z \in 
\left\lbrace -0.8,- 0.4,0.4,0.8\,\right\rbrace\,,
\right\rbrace$\\ 
\hline
\end{tabular}
\caption{The training data for learning the OPE coefficient $c_2$
when supplied the function $f(x)$ defined in Equation \eqref{eq:4ptCFT1 regression}.
We choose $N=100$ and the $\left\lbrace f(z_i)\right\rbrace$ are sampled at the $z_i$ defined
previously in Table 
\ref{tab:cbclassifier}.
The neural network was trained on 100000 instances of this data, i.e.
100000 functions with all modes $c_{n\leq 5}$ randomly sampled from
a normal distribution valued in $(0,1)$. The final value of $c_2$ was extracted from the regressor
by taking the median as described previously for the Fourier expansion.}
\label{tab:cb regressor}
\end{table}
As we saw for the classification example,
the neural network architecture that is needed here turns out to be somewhat simpler than in the Fourier case discussed
previously. In particular, it is a two layer network, consisting of 80 relu activated neurons in the first
layer, and 40 relu activated neurons in the second layer. This network was trained for 50 epochs, and the comparison
with actual values of $c_2$ and predicted values for 200 test functions is shown in Figure \ref{fig:CBRegressorValues}.
We again see that the neural network is able to predict the OPE coefficient $c_2$ appearing in the conformal block
expansion of putative four-point functions to good accuracy.
\subsection{Principal Components Analysis}
\label{subsection:Principal Components}
\begin{figure}[h!]
  \centering
      \includegraphics[width=0.8\linewidth]{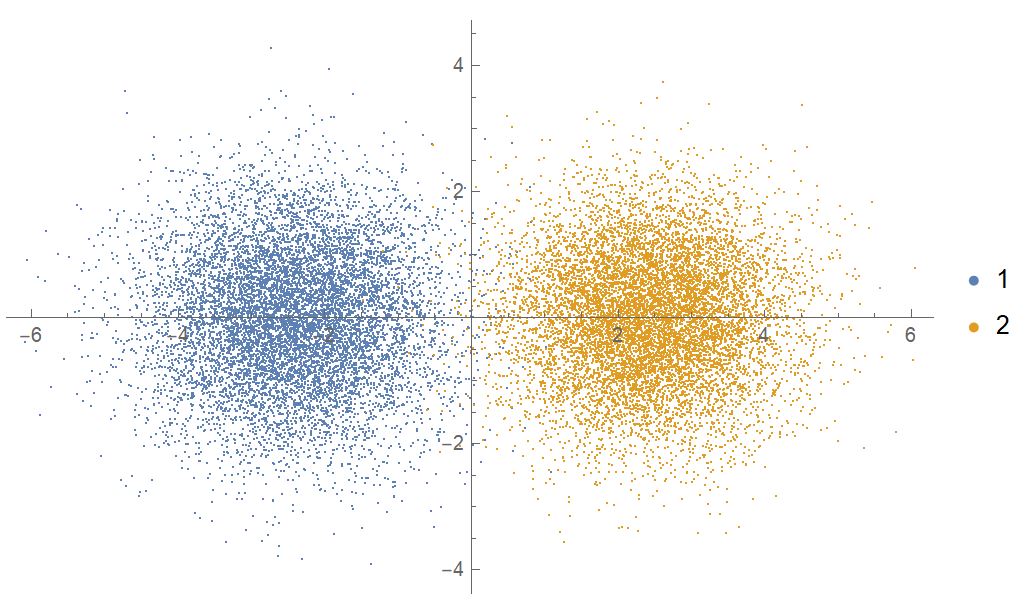}
  \caption{Principal component projections to 2-dimensions of the randomized
  conformal block expansions with 
  and without the $c_2$ term. These functions are defined respectively
  as $f^{(1)}(z)$ and $f^{(2)}(z)$ in Equation \eqref{eq:4ptCFT1 classification}.
  As before, $f^{(1)}(z)$s are represented by a 
  blue point cloud and $f^{(2)}(z)$s by a yellow point cloud.}
  \label{fig:pcaCB}
\end{figure}

\paragraph{}
The nature of this problems beckons for the distinction of a fundamental structure: since a simple neural network seems to tell whether a Fourier series or a conformal block expansion
contains a particular term, is there an underlying difference between the respective graphs of the functions?
We address this question first for the Fourier expansion and then for the conformal blocks.
To do so, it is useful to think of the data in Table \ref{tab:fourierclassifier} in terms of
point clouds in 100 dimensions as follows. If
\begin{equation}
    X := \left\{
- \frac{\pi}{2} + j \frac{\pi}{100}
\right\}_{j = 0, \ldots, 99}\,,
\end{equation}
is a list of 100 real points, then Table \ref{tab:fourierclassifier}  defines
sets of point-clouds in $\mathbb{R}^{100}$: 
the functions $f^{(1)}(x)$ and  $f^{(2)}(x)$ defined in \eqref{eq:fourier classes} and
evaluated at $X$ for random samples of coefficients $a_n$ drawn from a uniform distribution between $[0,1]$. Are these two data sets structurally different?
Figure \ref{fig:fourier classification} might suggest a disparity and the neural network of Figure \ref{fig:FourierClassifierNN} does distinguish the two.
\begin{figure}[h!]
  \centering
      \includegraphics[width=0.8\linewidth]{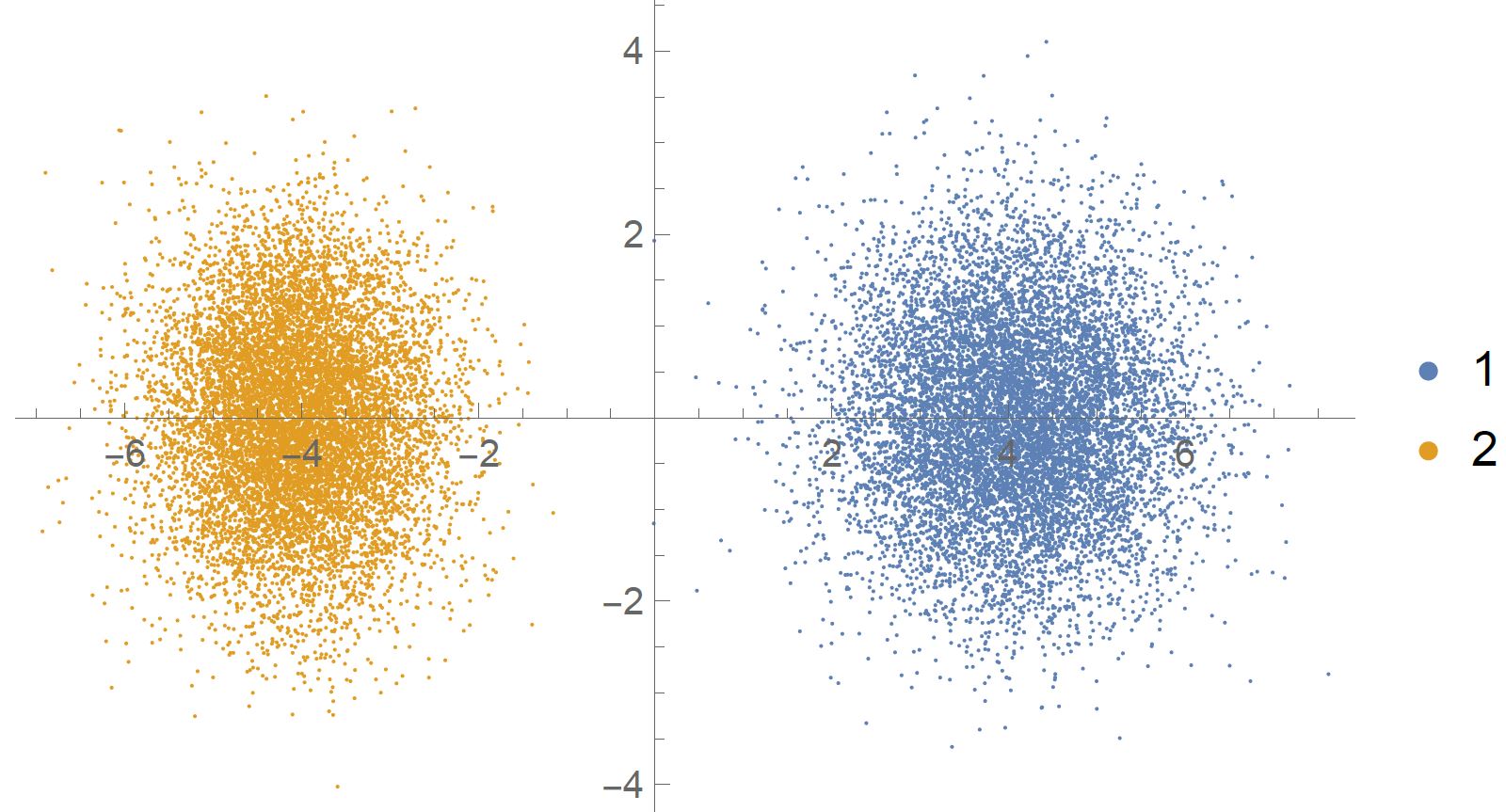}
  \caption{Principal component projections to 2-dimensions of the randomized 4-term Fourier expansions with 
  and without the $\sin(2x)$ term. These functions are defined respectively
  as $f^{(1)}(x)$ and $f^{(2)}(x)$ in Equation \eqref{eq:fourier classes}. The functions $f^{(1)}(x)$ appear
  as a blue point cloud and $f^{(2)}(x)$ as a yellow point cloud.}
  \label{fig:pcaFourier}
\end{figure}
Of course, 100-dimensions is not visualizable. Nevertheless, a standard technique lends itself. One can apply {\it principal component analysis} (PCA) where one can project down (while minimizing on the Euclidean distance to the lower dimension).
We apply PCA and reduce to 2-dimensions for ease to the eye, and this is shown in Figure \ref{fig:pcaFourier}, with 1 (blue) and 2 (orange) marking $f^{(1)}$ and $f^{(2)}$ respectively.
We see that the components are completely separated and thus indeed Fourier series with and without a particular term are different. One may similarly apply PCA on the conformal block classification problem of Section
\ref{subsection:cb}. The procedure is precisely the same as the Fourier case above. 
We then find that the datasets separate as per Figure \ref{fig:pcaCB} with the same
color conventions as in the Figure \ref{fig:pcaFourier}. 

\section{Neural Networks applied to the 3d Ising Model}\label{Section:3d-Ising}
\paragraph{}
In this section, we would like to use machine-learning to study the following questions.
Firstly, we would like to train a neural network algorithm to classify
operators in the 3d Ising Model based on their $\mathbb{Z}_2$ parity.
This is a natural classification problem, along the lines of the ones
addressed previously. Further, we shall investigate if
it is possible to train a neural network on the existing
spectral data \cite{Simmons-Duffin:2016wlq} 
to estimate the conformal 
dimension of an operator given its spin by means of regression.
We should emphasize that a key difference between conventional regression
methods and neural network regression is that in conventional methods we
fit a pre-selected curve, such as a straight line, or polynomial curve to
the given data, while the neural network is simply fed the data and its 
usual hyper-parameters, and `decides' the fitting curve on its own.

\subsection{The 3d Ising Model CFT: Review}\label{Subcsection:Ising Review}
\paragraph{}
We now collect some relevant basic details about the 3d Ising Model, as well
as a summary of our notations and conventions,
referring the reader to
the review
\cite{Poland:2018epd} for further details.
\paragraph{}
The 3d Ising model is a set of random spins $s_i=\pm 1$ in a cubic lattice in $\mathbb{R}^3$ with nearest neighbor interactions.
The partition function is
\begin{equation}
\mathcal{Z} = \sum_{\lbrace s_i\rbrace} \exp \left(-J\sum_{\langle ij\rangle} s_i \cdot s_i\right)\,.
\end{equation}
We note here that the lattice theory is invariant under the $\mathbb{Z}_2$ symmetry $s_i\mapsto -s_i$.
It is well known that this model exhibits a phase transition, and at its critical point, 
conformal symmetry emerges and is described by a CFT, which is what we 
are interested in. We note here that neural networks have previously been
trained to recognize this and other such phase transitions from random spin configurations
generated by Monte-Carlo methods 
\cite{carrasquilla2017machine,ch2017machine,broecker2017machine,JMLR:v18:17-527,d2020learning}.
Our interest is in the CFT occurring at
this critical point.
A useful microscopic realization of this CFT is in terms of a scalar field theory in three dimensions.
\begin{equation}
S = \int d^3x\,\left(\tfrac12 \left(\partial\sigma\right)^2
+\tfrac12\,m^2\sigma^2+\tfrac{1}{4!}\,\lambda\sigma^4\,\right)\,,
\end{equation}
where both $m^2$ and $\lambda$ describe relevant couplings. At a critical
value of the dimensionless ratio $m^2\over\lambda^2$, the long distance behavior 
of this model is described by the same CFT as the lattice model above.
\paragraph{}
Let us now turn to reviewing the physical data of the 3d Ising Model CFT as obtained by the 
bootstrap \cite{ElShowk:2012ht,El-Showk:2014dwa,Kos:2014bka,Simmons-Duffin:2015qma,Kos:2016ysd}. 
The procedure adopted is to consider the four-point functions
$\left\langle \sigma\sigma\sigma\sigma\right\rangle$,
$\left\langle \epsilon\epsilon\epsilon\epsilon\right\rangle$, 
$\left\langle \sigma\sigma\epsilon\epsilon\right\rangle$ and the operator
product expansions 
\begin{equation}\label{eq: Ising OPE}
\sigma \times \sigma = \sum_{\co} f_{\sigma\sigma\co}\,\co+\ldots\,,\quad
\epsilon \times \epsilon = \sum_{\co}
f_{\epsilon\epsilon\co}\,\co+\ldots\,,\quad
\sigma \times \epsilon = \sum_{\co} f_{\sigma\epsilon\co}\,\co+\ldots\,,
\end{equation}
where the `$\ldots$' denote the contributions from the descendants.
Requiring the unitarity and crossing symmetry for the above four-point functions 
constrains the conformal dimensions and OPE coefficients above.
In particular as obtained in  \cite{Kos:2016ysd}:
\begin{equation}
\begin{split}
\Delta_\sigma = 0.5181489(10) \,,&\qquad f_{\sigma\sigma\epsilon}=1.0518537(41)\,,\\
\Delta_\epsilon =1.412625(10)\,, &\qquad  f_{\epsilon\epsilon\epsilon}=1.532435(19)\,.
\end{split}
\end{equation}
The numerical and analytical results for the 
spectrum and the OPE coefficients of additional operators $\mathcal{O}$ required for crossing symmetry
of these correlators were obtained in \cite{Simmons-Duffin:2016wlq} for more than a hundred operators.
\paragraph{}
Here, we point out the existence of an important sub-class 
of operators known as \textit{double-twist operators} \cite{Fitzpatrick:2012yx,Komargodski:2012ek}. 
It has been proven for a CFT in dimensions greater than
two, and containing operators $\mathcal{O}_1$ and $\mathcal{O}_2$ of twist $\tau_1$ and $\tau_2$
that there exist infinite families of operators $[\mathcal{O}_1\mathcal{O}_1]_n$ of increasing spin and whose
twist\footnote{The twist $\tau$ of an operator equals $\Delta-\ell$, where $\Delta$ is the conformal dimension 
and $\ell$ is the spin of the operator.}
$\tau$ approaches $\tau_1+\tau_2+2n$ as for all integers $n\geq 0$. These played an important role in the analysis
of \cite{Simmons-Duffin:2016wlq}. 
We will shortly use the numerical results obtained there
to train a neural network regressor to predict the twists of operators in the
families $[\sigma\sigma]_0$ and $[\sigma\epsilon]_0$.

\paragraph{}
To sum up, the critical behavior of the 3d Ising model is described by
a CFT with a $\mathbb{Z}_2$ global symmetry and two relevant 
scalars, $\sigma$, which is
$\mathbb{Z}_2$-odd and $\epsilon$, which is
$\mathbb{Z}_2$ even.
Furthermore, physical data of the CFT, i.e. the conformal dimensions 
$\Delta_\sigma$ and $\Delta_\epsilon$
as well as the OPE coefficients $f_{\sigma\sigma\epsilon}$ and 
$f_{\epsilon\epsilon\epsilon}$ to lie in a tiny island by crossing 
symmetry of the scalar four-point function. Further, crossing symmetry
also determines the conformal dimensions and OPE coefficients of
operators appearing in the $\sigma\times \sigma$, $\sigma\times\epsilon$
and $\epsilon\times\epsilon$ operator product expansions (OPEs), though
the precise evaluation of these numbers is a formidable numerical
and analytical task \cite{Simmons-Duffin:2016wlq}.

\subsection{Learning The $\mathbb{Z}_2$ Symmetry of the 3d Ising Model}
\label{Subsection:Z2-Ising}
\paragraph{}
Consider the OPEs \eqref{eq: Ising OPE} of the 3d Ising Model.
As $\sigma$ is $\mathbb{Z}_2$-odd and $\epsilon$ is $\mathbb{Z}_2$-even,
the operators $\co$ appearing in the $\sigma\times\sigma$ and $\epsilon\times\epsilon$ OPEs are
$\mathbb{Z}_2$ even, while the operators appearing in the $\sigma\times\epsilon$
OPE are $\mathbb{Z}_2$-odd. That is, 
a $\mathbb{Z}_2$-odd operator $\co$ has a strictly zero OPE
coefficient $ f_{\sigma\sigma\co}$ and $f_{\epsilon\epsilon\co}$, and 
a non-zero OPE coefficient $f_{\sigma\epsilon\co}$,
while a  $\mathbb{Z}_2$-even operator has a non-zero OPE coefficient for
at least one of $ f_{\sigma\sigma\co}$ and $f_{\epsilon\epsilon\co}$, 
and a strictly zero OPE coefficient
$f_{\sigma\epsilon\co}$. 

\begin{figure}[h!]
  \centering
  \begin{subfigure}[b]{0.45\linewidth}
    \includegraphics[width=\linewidth]{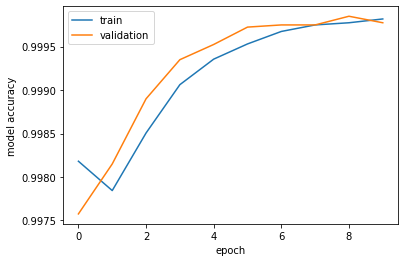}
    \caption{Accuracy}
  \end{subfigure}
  \begin{subfigure}[b]{0.45\linewidth}
    \includegraphics[width=\linewidth]{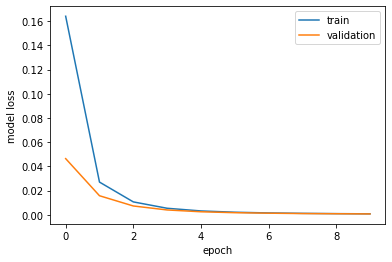}
    \caption{Loss}
  \end{subfigure}
  \caption{The Accuracy and Loss curves for the $\mathbb{Z}_2$
  classification Neural Network. The architecture of this network
  is outlined in Figure \ref{fig:Z2IsingClassifier}. }
  \label{fig:Z2Ising}
\end{figure}
\begin{figure}
  \centering
      \includegraphics[width=0.5\linewidth]{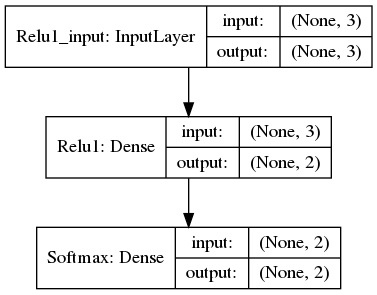}
  \caption{The Neural Network Classifier for classifying operators under a
  $\mathbb{Z}_2$ symmetry in a CFT. This
  has 68 trainable parameters.}
  \label{fig:Z2IsingClassifier}
\end{figure}

\paragraph{}
Here we shall train a neural network to classify operators
that appear in the OPE of $\sigma\times\sigma$, $\sigma\times\epsilon$
and $\epsilon\times\epsilon$
under this $\mathbb{Z}_2$ symmetry. The training data for
this classification is summarized in Table \ref{tab:Z2Ising}, and
the architecture of the neural network classifier is shown in Figure 
\ref{fig:Z2IsingClassifier}.
We achieved 99.85 percent accuracy by training for 10 epochs.
\begin{table}
\centering
\begin{tabular}{|c||c|c|c|c|}
\hline
$\mathbb{Z}_{2}$ charge 
&$\mathrm{f}_{\sigma\sigma\mathcal{O}}$ 
& ${f}_{\epsilon\epsilon\mathcal{O}}$ 
& ${f}_{\sigma\epsilon\mathcal{O}}$
& $\mathrm{Y}$ 
\\ \hline
$+1$& $x_1$ & $x_2$ & $0$ & $\left(1,0\right)$\\ 
$-1$&0 & 0 & $x_3$ & $\left(0,1\right)$\\
\hline
\end{tabular}
\caption{The training data for the classification
of CFT operators $\mathcal{O}$ under a putative $\mathbb{Z}_2$ symmetry 
of the kind that appears in the 3d Ising Model. We continue to label
the two fundamental fields of the CFT as $\sigma$ and $\epsilon$, and
again $\sigma$ is taken to have $\mathbb{Z}_2$ charge $-1$ and 
$\epsilon$ is taken to have $\mathbb{Z}_2$ charge $+1$. 
The left-most column indicates the $\mathbb{Z}_2$ charge
of $\mathcal{O}$ and is not part of the training data.
The numerical 
values of the $x_i$s are randomly chosen positive real numbers between 
0 and 1. We trained the classifier on on 100,000 instances of such data.}
\label{tab:Z2Ising}
\end{table}
Having trained this classifier, we would like to examine how well it performs
on the actual OPE data of the 3d Ising model obtained in \cite{Simmons-Duffin:2016wlq},
see their Tables 1 to 7.
Naively feeding this data to the classifier yields indifferent results. We believe that
this was so because the OPE coefficients that appear in the 3d Ising model
take numerical values ranging from $\mathcal{O}(1)$ to $\mathcal{O}(10^{-12})$ or so, while the classifier was
trained on $\mathcal{O}(1)$ values. Indeed, once we transform the OPE data 
to take on positive values of order 1, \textit{via} the 
function $f_{transformed}=\vert \log \vert f_{original} \vert \vert$, we find that
the classifier works to 100 percent accuracy with no mis-classifications
\footnote{One may try to train the classifier on a distribution of data that range
that the actual Ising OPE coefficients lie in, but this leads to a degraded performance
during training itself. This appears to be a particular instance of the general
feature that neural networks train and perform poorly on data with extreme
ranges and 
distributions. Also, though we have not done so here, we would expect that a
comparably good way or processing the OPE coefficients would be to normalize
them by their mean field values, which also results in $\mathcal{O}(1)$ numbers.}.

\subsection{Classifying higher rank discrete symmetries}\label{Subsection:Higher Discrete Symmetry}
\begin{figure}
  \centering
      \includegraphics[width=0.5\linewidth]{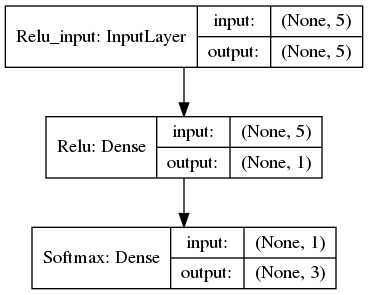}
  \caption{The Neural Network Classifier for classifying operators under a
  $\mathbb{Z}_3$ symmetry in a CFT. This
  has 12 trainable parameters.}
  \label{fig:Z3ClassifierNN}
\end{figure}

\begin{figure}
  \centering
  \begin{subfigure}[b]{0.45\linewidth}
    \includegraphics[width=\linewidth]{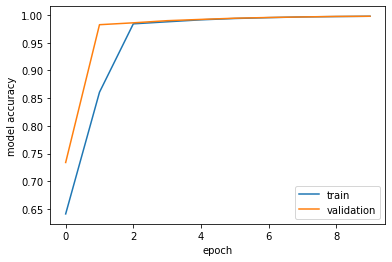}
    \caption{Accuracy}
  \end{subfigure}
  \begin{subfigure}[b]{0.45\linewidth}
    \includegraphics[width=\linewidth]{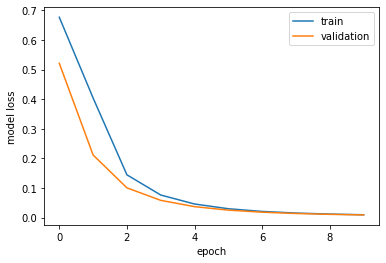}
    \caption{Loss}
  \end{subfigure}
  \caption{The Accuracy and Loss curves for the $\mathbb{Z}_3$ Classification.}
  \label{fig:Z3Curves}
\end{figure}

\paragraph{}
It is also possible to carry out this classification with higher rank discrete
symmetries. We consider a putative
a CFT where there exist scalar fields $\varphi_{i}$ that
carry a representation of $\mathbb{Z}_3$. We label the fields
$\varphi_{0}$, $\varphi_{1}$ and $\varphi_{2}$, where the subscripts are the  $\mathbb{Z}_3$ charges of the 
respective operators.
Operator product expansions in this CFT would take the form
\begin{equation}\label{eq:z3 symm}
\begin{split}
\varphi_{0} \times \varphi_{0} &= \sum_{\co_{(0)}} f_{0\,0\,\co_{(0)}}\,\co_{(0)}+\ldots\,,\qquad
\varphi_{0} \times \varphi_{1} = \sum_{\co_{(1)}} f_{0\,1\,\co_{(1)}}\,\co_{(1)}+\ldots\,,\\
\varphi_{0} \times \varphi_{2} &= \sum_{\co_{(2)}} f_{0\,2\,\co_{(2)}}\,\co_{(2)+}+\ldots\,,\qquad
\varphi_{1} \times \varphi_{2} = \sum_{\co_{(0)}} f_{1\,2\,\co_{(0)}}\,\co_{(0)}+\ldots\,,
\end{split}
\end{equation}
where as usual the `$\ldots$' denotes the contributions from the descendants.
The operator $\co_{(i)}$ has charge $i$ under the $\mathbb{Z}_3$ symmetry. 
Operators $\co_{(0)}$ would have generically non-zero 
OPE coefficients $f_{\varphi_0\,\varphi_0\,\co_{(0)}}$ 
and $f_{\varphi_1\,\varphi_2\,\co_{(0)}}$, operators $\co_{(1)}$ 
would have a non-zero $f_{\varphi_0\,\varphi_0\,\co_{(1)}}$ and lastly,
operators $\co_{(2)}$ would have generically non-zero OPE coefficients
$f_{\varphi_1,\varphi_1\,\co_{(2)}}$  and $f_{\varphi_0\,\varphi_0\,\co_{(2)}}$ 
respectively. All other OPE coefficients would be zero on account of the $\mathbb{Z}_3$
symmetry. 
\paragraph{}
We generated the data representing these features as per Table
\ref{tab:Z3 symm}, and trained the neural network classifier described in Figure
\ref{fig:Z3ClassifierNN} on it.
\begin{table}
\centering
\begin{tabular}{|c||c|c|c|c|c|c|}
\hline
$\mathbb{Z}_{3}$ charge 
&$\mathrm{f}_{\varphi_{0}\varphi_{0}\mathcal{O}}$ 
& ${f}_{\varphi_{0}\varphi_{1}\mathcal{O}}$ 
& ${f}_{\varphi_{0}\varphi_{2}\mathcal{O}}$
& ${f}_{\varphi_{1}\varphi_{1}\mathcal{O}}$
& ${f}_{\varphi_{1}\varphi_{2}\mathcal{O}}$
& $\mathrm{Y}$ 
\\ \hline
$0$& $x_1$ & $0$ & $0$ & 0& $x_2$& $\left(1,0,0\right)$\\ 
$1$&0 & $x_3$ & 0 & 0& 0 &$\left(0,1,0\right)$\\
$2$&0 & 0 & $x_4$ & $x_5$ & 0&$\left(0,0,1\right)$\\
\hline
\end{tabular}
\caption{The training data for the classification
of CFT operators $\mathcal{O}$ under a putative $\mathbb{Z}_3$ symmetry 
of the kind defined in Equation \ref{eq:z3 symm}. 
The left-most column indicates the $\mathbb{Z}_3$ charge
of $\mathcal{O}$ and is not part of the training data.
The numerical 
values of the $x_i$s are randomly chosen positive real numbers between 
0 and 1. We trained the classifier on on 100,000 instances of such data.}
\label{tab:Z3 symm}
\end{table}
The accuracy and loss curves during training are shown in Figure
\ref{fig:Z3Curves}. We could train to 99.78 percent accuracy over 10 epochs.
We also tested the classifier on 2000 instances of freshly generated putative OPE data of the kind detailed in Table \ref{tab:Z3 symm}, which it had not been trained 
on. It could classify those data to 99.73 percent accuracy.
We note here that it should be possible to similarly consider 
higher-rank symmetry groups like $\mathbb{Z}_4$ and $\mathbb{Z}_2\times\mathbb{Z}_2$
and especially $\mathrm{S}_N\times\mathbb{Z}_2$ and try to train 
neural network classifiers to recognize these symmetries from the
OPE coefficients. The CFTs with such discrete symmetries have been studied via
conformal bootstrap in \cite{Rong:2017cow} and \cite{Stergiou:2018gjj}. It would be
interesting to examine if the operators there can be similarly classified 
under their respective discrete symmetries by
these ML methods.

\subsection{Learning the Spectrum of the 3d Ising Model}\label{Subsection:3d Ising Spectrum}
\begin{figure}
  \centering
  \begin{subfigure}[b]{0.45\linewidth}
    \includegraphics[width=\linewidth]{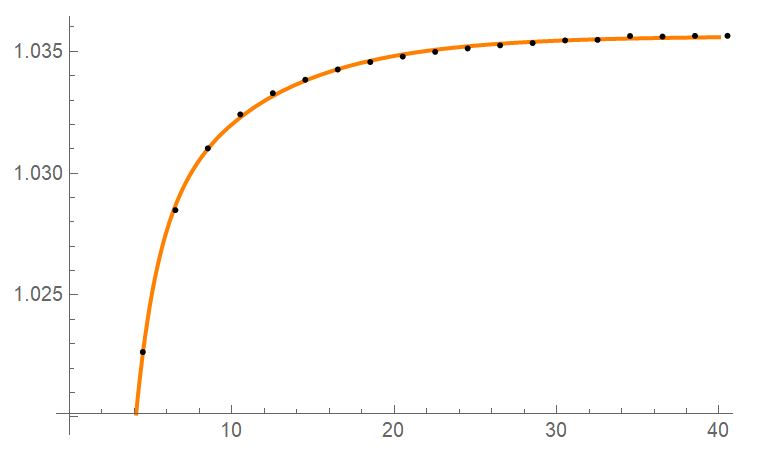}
    \caption{Regression for $\tau_{[\sigma\epsilon]_0}$}
  \end{subfigure}
  \begin{subfigure}[b]{0.45\linewidth}
    \includegraphics[width=\linewidth]{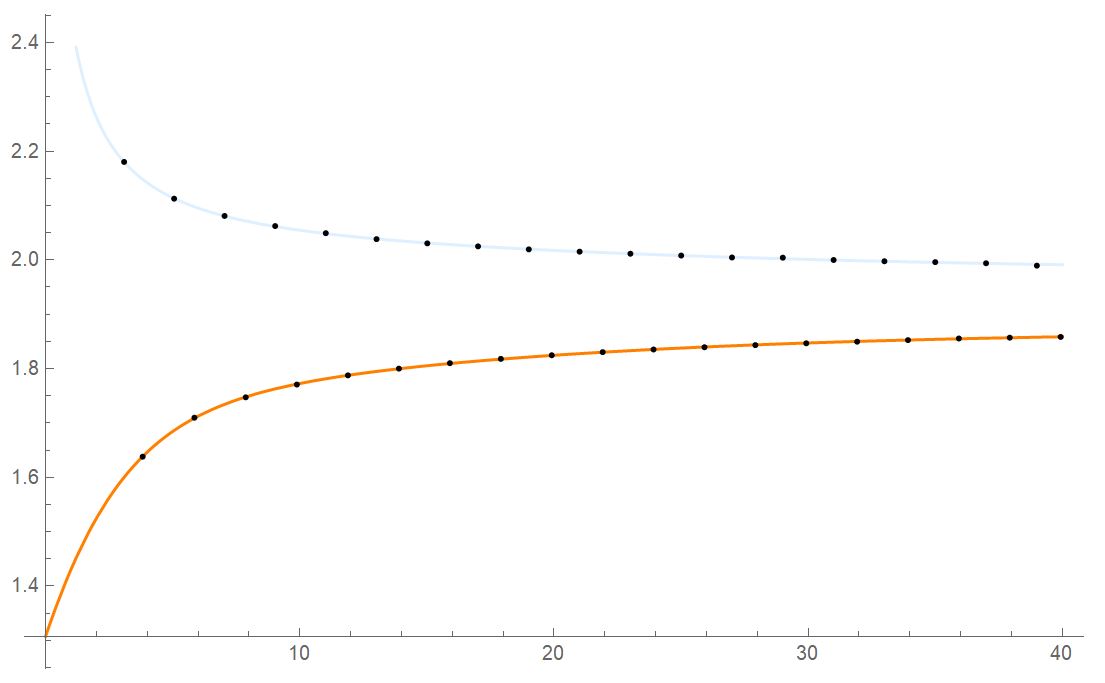}
    \caption{Regression for $\tau_{[\sigma\sigma]_0}$}
  \end{subfigure}
  \caption{Directly fitting a functional form to the 3d Ising data. We see that a judicious but somewhat non-obvious
  choice of fitting functions \eqref{eq:tanh ansatz} leads to an impressive fit with the available data.}
  \label{fig:tanhCurves}
\end{figure}

\begin{figure}
  \centering
      \includegraphics[width=0.5\linewidth]{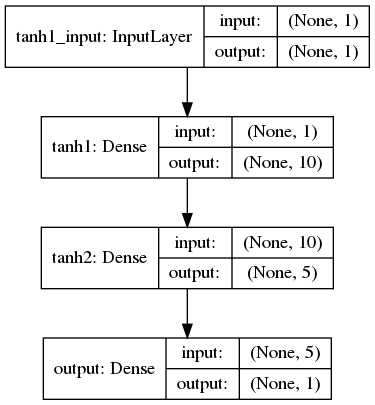}
  \caption{The Neural Network Regressor for predicting operator twists in the 
  $[\sigma\sigma]_0$ and $[\sigma\epsilon]_0$ families in the 3d Ising Model. This
  has 81 trainable parameters and is run for 6000 epochs. The corresponding regression curves
  appear in Figure }
  \label{fig:IsingRegressorTanh}
\end{figure}

\begin{figure}
  \centering
  \begin{subfigure}[b]{0.48\linewidth}
    \includegraphics[width=\linewidth]{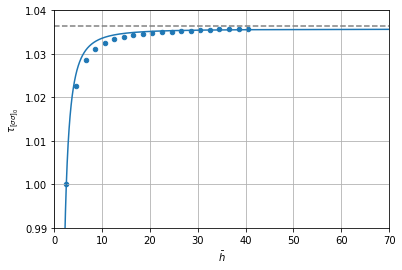}
    \caption{Regression for $\tau_{[\sigma\sigma]_0}$}
  \end{subfigure}
  \begin{subfigure}[b]{0.48\linewidth}
    \includegraphics[width=\linewidth]{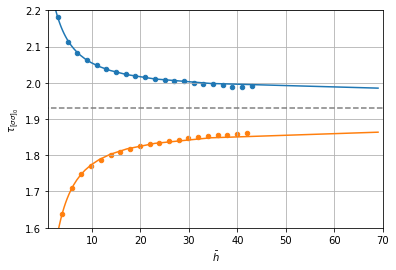}
    \caption{Regression for $\tau_{[\sigma\epsilon]_0}$}
  \end{subfigure}
  \caption{Neural Network Regression for the 3d Ising Model spectrum. The smooth lines are the regression curves, the points are the numerical values obtained in \cite{Simmons-Duffin:2016wlq} without the error bars.
  The dotted line in (a) denotes the asymptotic value $\tau=2\Delta_\sigma$, while the dotted line in (b)
  denotes the asymptotic value $\tau=\Delta_\sigma+\Delta_\epsilon$. The curves are extrapolated out to
  $\bar{h}=70$. The r-squared coefficient for (a) is 0.9804, and for (b) is 0.9966 for the even spin operators
  (blue line) and 0.9926 for the odd spin operators (orange line).}
  \label{fig:IsingRegressionCurves}
\end{figure}

\paragraph{}
Having examined the performance of neural networks on various classification tasks,
we now turn to regression tasks. In particular, we shall train a neural network 
on the available spectral data of the 3d Ising model obtained in
\cite{Simmons-Duffin:2016wlq} by combining numerical and analytical bootstrap
methods to learn the twist of the exchange operator as a function
of its spin. We shall also further specialize to the OPE families $[\sigma\sigma]_0$
and $[\sigma\epsilon]_0$. The fitting data for these operators are taken from 
Tables 3 and 6 of \cite{Simmons-Duffin:2016wlq}. We remind the reader that these
tables give the data for the quantum numbers $\tau$ and $\ell$, which are related
to $\bar{h}$ \textit{via}
\begin{equation}
    \bar{h}=\frac{\tau}{2}+\ell\,.
\end{equation}
Also, the twists of operators of even spin and odd spin fall into two different curves in the $[\sigma\epsilon]_0$
family. It is interesting to note at this point that the three curves can also be fit almost exactly with a judicious
choice of fitting ansatz. In particular, we consider the functions
\begin{equation}\label{eq:tanh ansatz}
\begin{split}
    \tau_1\left(\bar{h}\right) &= \tanh\left(a_1\,\bar{h}+b_1\right)+ \tanh\left(a_2\,\bar{h}+b_2\right)+c\,,
    \\
    \tau_2\left(\bar{h}\right) &= \frac{p}{\tanh\left(q_1\,\bar{h}+r_1\right)-\tanh\left(q_2\,\bar{h}+r_2\right)}+s\,.
\end{split}
\end{equation}
The data for $\tau_{[\sigma\sigma]_{(0)}}$ turns out to fit to $\tau_1$ 
with the parameters
\begin{equation}
    a_1=0.0740199\,,\,b_1= 2.42646\,,\, a_2=0.3991\,,\,b_2= 1.18916\,,\, c=-0.964377\,,
\end{equation}
with R-squared $1.0$ and total (corrected) squared error less than $0.001$.
Likewise, the data for odd spin operators in the $[\sigma\sigma]_{(0)}$ family fits to $\tau_1$ 
with the parameters
\begin{equation}
    a_1=0.0349198\,,\,b_1= 1.18396\,,\, a_2=0.203412\,,\, b_2= 0.706744\,,\, c=-0.129985\,,
\end{equation}
while the odd spin operators fit to $\tau_2$ with the parameters
\begin{equation}
\begin{split}
    p&= 0.024275\,,\,q_1 =0.0767471\,,\, r_1=1.13926\,,\\q_2 &= -0.0242292\,,\,r_2=1.09043\,,
    s=1.96356\,,
\end{split}
\end{equation}
also with R-squared around $1.0$. The corresponding regression curves are plotted in Figure \ref{fig:tanhCurves}.
It is perhaps curious that the relatively simple functional forms of Equation \eqref{eq:tanh ansatz} seem to 
effectively capture the much more complicated expressions obtained through direct analytical methods
\cite{Simmons-Duffin:2016wlq}, and it may be interesting to understand this further.
For the moment, we will simply note that the success of this ansatz also suggests that we may try training a neural
network on this data with $\tanh$ activation functions. We find that the size of the network required to produce
comparable results is dramatically smaller than when choosing say sigmoid or
relu activation functions, reducing to a layer of 15 neurons followed by a second hidden layer of 10 neurons. 
This network is displayed in Figure \ref{fig:IsingRegressorTanh}.
Further, when supplying the data to the neural network for fitting, we found that
performance improved when the data was fed in the form
\begin{equation}
    \left(X,Y\right) \qquad :\qquad \left(\bar{h}^{-1},\tau\right)\,,
\end{equation}
for $\tau_{[\sigma\sigma]_0}$ regression and
\begin{equation}
    \left(X,Y\right) \qquad :\qquad \left(\bar{h}^{-2},\tau\right)\,,
\end{equation}
for $\tau_{[\sigma\epsilon]_0}$ regression. The training of the regressor was significantly more
stable when the data was fed in this form. 
The regression curves obtained after training for six thousand epochs are shown in Figure
\ref{fig:IsingRegressionCurves}, along with their extrapolations to $\bar{h}=70$.
\paragraph{}
Before we close this section we would like to make a few comments about the choice of activation function. We have
remarked previously that an important feature of ML based regression methods is that we do not feed the network
with a trial form of the function to fit to. Why, then, is the architecture of the regressor
so sensitive to the choice of activation
function? Or at least, the architecture seems indifferent to choosing between relu and sigmoid activations but 
simplifies dramatically when the tanh activation function is chosen. We believe that at least some 
insight might be gained from
the fitting functions in Equation \eqref{eq:tanh ansatz}. These suggest that the tanh functions are a 
particularly nice basis choice to expand these curves in, and the choice of activation function is at least
intuitively a choice of basis functions to expand the neural network decision function in. A `good' basis 
would have fewer undetermined coefficients, and this is what we seem to have found somewhat serendipitously
with our choice
of tanh as the activation function.

\section*{Acknowledgements}
The work of HYC was supported in
part by Ministry of Science and Technology through the grant 108-2112-M-002-004.
YHH would like to thank UK STFC grant 
ST/J00037X/1, 
as well as Minhyong Kim for conversations on symmetries from modular arithmetic.
SL would like to thank Pedro Leal for several helpful discussions on machine learning and 
neural networks.
SL’s work is supported by the Simons Foundation grant 488637 (Simons Collaboration
on the Non-perturbative bootstrap) and the project CERN/FIS-PAR/0019/2017. Centro
de Fisica do Porto is partially funded by the Foundation for Science and Technology of
Portugal (FCT) under the grant UID-04650-FCUP. The neural networks used here were programmed
using Keras and the notebooks of \cite{MEHTA20191} were helpful in structuring some of SL's code.

\appendix
\section*{Appendices}
In these appendices, we briefly summarize some rudiments on neural networks and principal components which are used in the text.
\section{Neural Networks}
\paragraph{}
In this section, we provide a quick introduction to neural networks. This section is far from an exhaustive review on the working of neural networks, we only aim to lay down the relevant terms and a few notations. Neural networks are extremely useful for performing machine learning tasks such as classification and regression. A typical neural network consists of neurons also referred to as the single layer perceptron (\emph{SLP}), arranged in layers on top of each other. Each such neuron is for all practical purposes a function of a vector valued input $z$, with components $z_{i}$ and is arranged to have in it real valued parameters $\omega_{i}$ called weights and $b$ called the bias. Thus, an \emph{SLP} is a function of the form $f(\sum_{i}\omega_{i}x_{i}^{j}+b)$. Where $x_{i}^{j}$ is the $i$th component of the $j$th input. The function $f(z)$ familiar to neural network practitioners as the \emph{activation function} is often taken to be a sigmoid, 
\begin{equation}
 f(z) = \sigma(z)= \frac{1}{1+exp(-z)}\,,
\end{equation}
though other choices such as $f(z)=\tanh(z)$ and the rectifier function
\begin{equation}
f(z)=\max\left(0,z\right)\,,
\end{equation}
are also equally popular. A neuron activated by the rectifier function is known as a rectifier linear unit,
or a relu. We have used relus almost as a \textit{de facto} choice in our neural networks above.
\paragraph{}
The process of training a neural network is then to find appropriate values for the weights $w_{i}$ and biases $b$ such that the desired goal is achieved. This is illustrated by the following example, suppose we are to perform a classification task using neural networks and we are supplied with a labeled dataset of the form D $= \{x_{i}^{j}, d^{j}\}$, where $x_{i}$ is some vector belonging to a class specified by the labels $d^{j}$. The task is to train a neural network to \emph{accurately} predict the class label of an instance not contained in D. This task can be achieved in two steps, first by specifying what is known to the neural network practitioners as the \emph{loss function}. For classification tasks a typical loss function is the following, 
\begin{equation}
 f_{loss}= \sum_{j}(f(\omega_{i}x_{i}^{j}+b)-d^{j})^{2}.
\end{equation}
The second step is to optimize the loss function over the parameters $w_{i}$ and $b$ using algorithms such as the \emph{Stochastic gradient descent}. The performance of a neural network after training is evaluated on a part of the dataset D set aside before training. This is known as the \emph{test data}, while D modulo \emph{test data} $=$ \emph{training data}. The performance of a neural network is evaluated with respect to some metric, typically in a classification task, the metric is \emph{accuracy}. Thus the essentials associated to a neural network are, The training data, the test data, the activation functions, the loss function and the optimization algorithm.

\section{Principal Component Analysis (PCA)}
\paragraph{}
In this section we briefly introduce the principal component analysis technique, which is employed for creating a low dimensional representation for a very high dimensional dataset. Again, we do not attempt an exhaustive review. We will give a brief illustration of a situation where PCA might come handy. 
\paragraph{}
Suppose we are given a dataset D $\{x_{i}^{j}, d^{j}\}$ for a classification task. Let's say that the vectors $x^{j}$ belong to a very high dimensional space of dimensionality $n$ and it may be desirable to use low dimensional projections of   $x^{j}$s for performing the classification. However the projections must be such that information from all the original features is retained. This is precisely what PCA achieves. This is done by defining new variables or features $y_{i}^{j}$ such that,
\begin{equation}
 y_{i}^{j} = f_{i}^{j}(x_{1},x_{2},\cdots, x_{n}).
\end{equation}
These new variables called principal components are chosen such that they are orthogonal to each other. After the principal components are defined, dimensionality reduction is achieved by keeping only those components which are of most relevance, the relevance being often measured by the \emph{variance explained}.

\bibliographystyle{JHEP}
\bibliography{refs}

\providecommand{\href}[2]{#2}\begingroup\raggedright\begin{thebibliography}{10}

\bibitem{Polyakov:1988md}
A.~M. Polyakov, \emph{{Fermi-Bose Transmutations Induced by Gauge Fields}},
  \href{https://doi.org/10.1142/S0217732388000398}{\emph{Mod. Phys. Lett. A}
  {\bfseries 3} (1988) 325}.

\bibitem{Shaji:1990is}
N.~Shaji, R.~Shankar and M.~Sivakumar, \emph{{On Bose-fermi Equivalence in a
  U(1) Gauge Theory With {Chern-Simons} Action}},
  \href{https://doi.org/10.1142/S0217732390000664}{\emph{Mod. Phys. Lett. A}
  {\bfseries 5} (1990) 593}.

\bibitem{Fradkin:1994tt}
E.~H. Fradkin and F.~A. Schaposnik, \emph{{The Fermion - boson mapping in
  three-dimensional quantum field theory}},
  \href{https://doi.org/10.1016/0370-2693(94)91374-9}{\emph{Phys. Lett. B}
  {\bfseries 338} (1994) 253}
  [\href{https://arxiv.org/abs/hep-th/9407182}{{\ttfamily hep-th/9407182}}].

\bibitem{Giombi:2011kc}
S.~Giombi, S.~Minwalla, S.~Prakash, S.~P. Trivedi, S.~R. Wadia and X.~Yin,
  \emph{{Chern-Simons Theory with Vector Fermion Matter}},
  \href{https://doi.org/10.1140/epjc/s10052-012-2112-0}{\emph{Eur. Phys. J. C}
  {\bfseries 72} (2012) 2112}
  [\href{https://arxiv.org/abs/1110.4386}{{\ttfamily 1110.4386}}].

\bibitem{Aharony:2011jz}
O.~Aharony, G.~Gur-Ari and R.~Yacoby, \emph{{d=3 Bosonic Vector Models Coupled
  to Chern-Simons Gauge Theories}},
  \href{https://doi.org/10.1007/JHEP03(2012)037}{\emph{JHEP} {\bfseries 03}
  (2012) 037} [\href{https://arxiv.org/abs/1110.4382}{{\ttfamily 1110.4382}}].

\bibitem{Aharony:2012nh}
O.~Aharony, G.~Gur-Ari and R.~Yacoby, \emph{{Correlation Functions of Large N
  Chern-Simons-Matter Theories and Bosonization in Three Dimensions}},
  \href{https://doi.org/10.1007/JHEP12(2012)028}{\emph{JHEP} {\bfseries 12}
  (2012) 028} [\href{https://arxiv.org/abs/1207.4593}{{\ttfamily 1207.4593}}].

\bibitem{Klebanov:2002ja}
I.~Klebanov and A.~Polyakov, \emph{{AdS dual of the critical O(N) vector
  model}}, \href{https://doi.org/10.1016/S0370-2693(02)02980-5}{\emph{Phys.
  Lett. B} {\bfseries 550} (2002) 213}
  [\href{https://arxiv.org/abs/hep-th/0210114}{{\ttfamily hep-th/0210114}}].

\bibitem{Sezgin:2002rt}
E.~Sezgin and P.~Sundell, \emph{{Massless higher spins and holography}},
  \href{https://doi.org/10.1016/S0550-3213(02)00739-3}{\emph{Nucl. Phys. B}
  {\bfseries 644} (2002) 303}
  [\href{https://arxiv.org/abs/hep-th/0205131}{{\ttfamily hep-th/0205131}}].

\bibitem{Gaberdiel:2010pz}
M.~R. Gaberdiel and R.~Gopakumar, \emph{{An AdS\_3 Dual for Minimal Model
  CFTs}}, \href{https://doi.org/10.1103/PhysRevD.83.066007}{\emph{Phys. Rev. D}
  {\bfseries 83} (2011) 066007}
  [\href{https://arxiv.org/abs/1011.2986}{{\ttfamily 1011.2986}}].

\bibitem{Beccaria:2014zma}
M.~Beccaria and A.~A. Tseytlin, \emph{{Vectorial AdS$_5$/CFT$_4$ duality for
  spin-one boundary theory}},
  \href{https://doi.org/10.1088/1751-8113/47/49/492001}{\emph{J. Phys. A}
  {\bfseries 47} (2014) 492001}
  [\href{https://arxiv.org/abs/1410.4457}{{\ttfamily 1410.4457}}].

\bibitem{Bae:2016hfy}
J.-B. Bae, E.~Joung and S.~Lal, \emph{{On the Holography of Free Yang-Mills}},
  \href{https://doi.org/10.1007/JHEP10(2016)074}{\emph{JHEP} {\bfseries 10}
  (2016) 074} [\href{https://arxiv.org/abs/1607.07651}{{\ttfamily
  1607.07651}}].

\bibitem{Rychkov:2011et}
S.~Rychkov, \emph{{Conformal Bootstrap in Three Dimensions?}},
  \href{https://arxiv.org/abs/1111.2115}{{\ttfamily 1111.2115}}.

\bibitem{ElShowk:2012ht}
S.~El-Showk, M.~F. Paulos, D.~Poland, S.~Rychkov, D.~Simmons-Duffin and
  A.~Vichi, \emph{{Solving the 3D Ising Model with the Conformal Bootstrap}},
  \href{https://doi.org/10.1103/PhysRevD.86.025022}{\emph{Phys. Rev. D}
  {\bfseries 86} (2012) 025022}
  [\href{https://arxiv.org/abs/1203.6064}{{\ttfamily 1203.6064}}].

\bibitem{El-Showk:2014dwa}
S.~El-Showk, M.~F. Paulos, D.~Poland, S.~Rychkov, D.~Simmons-Duffin and
  A.~Vichi, \emph{{Solving the 3d Ising Model with the Conformal Bootstrap II.
  c-Minimization and Precise Critical Exponents}},
  \href{https://doi.org/10.1007/s10955-014-1042-7}{\emph{J. Stat. Phys.}
  {\bfseries 157} (2014) 869}
  [\href{https://arxiv.org/abs/1403.4545}{{\ttfamily 1403.4545}}].

\bibitem{Kos:2016ysd}
F.~Kos, D.~Poland, D.~Simmons-Duffin and A.~Vichi, \emph{{Precision Islands in
  the Ising and $O(N)$ Models}},
  \href{https://doi.org/10.1007/JHEP08(2016)036}{\emph{JHEP} {\bfseries 08}
  (2016) 036} [\href{https://arxiv.org/abs/1603.04436}{{\ttfamily
  1603.04436}}].

\bibitem{Simmons-Duffin:2016wlq}
D.~Simmons-Duffin, \emph{{The Lightcone Bootstrap and the Spectrum of the 3d
  Ising CFT}}, \href{https://doi.org/10.1007/JHEP03(2017)086}{\emph{JHEP}
  {\bfseries 03} (2017) 086}
  [\href{https://arxiv.org/abs/1612.08471}{{\ttfamily 1612.08471}}].

\bibitem{Chester:2019ifh}
S.~M. Chester, W.~Landry, J.~Liu, D.~Poland, D.~Simmons-Duffin, N.~Su et~al.,
  \emph{{Carving out OPE space and precise $O(2)$ model critical exponents}},
  \href{https://arxiv.org/abs/1912.03324}{{\ttfamily 1912.03324}}.

\bibitem{Rong:2017cow}
J.~Rong and N.~Su, \emph{{Scalar CFTs and Their Large N Limits}},
  \href{https://doi.org/10.1007/JHEP09(2018)103}{\emph{JHEP} {\bfseries 09}
  (2018) 103} [\href{https://arxiv.org/abs/1712.00985}{{\ttfamily
  1712.00985}}].

\bibitem{Stergiou:2018gjj}
A.~Stergiou, \emph{{Bootstrapping hypercubic and hypertetrahedral theories in
  three dimensions}},
  \href{https://doi.org/10.1007/JHEP05(2018)035}{\emph{JHEP} {\bfseries 05}
  (2018) 035} [\href{https://arxiv.org/abs/1801.07127}{{\ttfamily
  1801.07127}}].

\bibitem{Beem:2014zpa}
C.~Beem, M.~Lemos, P.~Liendo, L.~Rastelli and B.~C. van Rees, \emph{{The $
  \mathcal{N}=2 $ superconformal bootstrap}},
  \href{https://doi.org/10.1007/JHEP03(2016)183}{\emph{JHEP} {\bfseries 03}
  (2016) 183} [\href{https://arxiv.org/abs/1412.7541}{{\ttfamily 1412.7541}}].

\bibitem{Beem:2015aoa}
C.~Beem, M.~Lemos, L.~Rastelli and B.~C. van Rees, \emph{{The (2, 0)
  superconformal bootstrap}},
  \href{https://doi.org/10.1103/PhysRevD.93.025016}{\emph{Phys. Rev. D}
  {\bfseries 93} (2016) 025016}
  [\href{https://arxiv.org/abs/1507.05637}{{\ttfamily 1507.05637}}].

\bibitem{Alday:2016njk}
L.~F. Alday, \emph{{Large Spin Perturbation Theory for Conformal Field
  Theories}}, \href{https://doi.org/10.1103/PhysRevLett.119.111601}{\emph{Phys.
  Rev. Lett.} {\bfseries 119} (2017) 111601}
  [\href{https://arxiv.org/abs/1611.01500}{{\ttfamily 1611.01500}}].

\bibitem{Gopakumar:2016cpb}
R.~Gopakumar, A.~Kaviraj, K.~Sen and A.~Sinha, \emph{{A Mellin space approach
  to the conformal bootstrap}},
  \href{https://doi.org/10.1007/JHEP05(2017)027}{\emph{JHEP} {\bfseries 05}
  (2017) 027} [\href{https://arxiv.org/abs/1611.08407}{{\ttfamily
  1611.08407}}].

\bibitem{Alday:2017zzv}
L.~F. Alday, J.~Henriksson and M.~van Loon, \emph{{Taming the
  $\epsilon$-expansion with large spin perturbation theory}},
  \href{https://doi.org/10.1007/JHEP07(2018)131}{\emph{JHEP} {\bfseries 07}
  (2018) 131} [\href{https://arxiv.org/abs/1712.02314}{{\ttfamily
  1712.02314}}].

\bibitem{Aharony:2018npf}
O.~Aharony, L.~F. Alday, A.~Bissi and R.~Yacoby, \emph{{The Analytic Bootstrap
  for Large $N$ Chern-Simons Vector Models}},
  \href{https://doi.org/10.1007/JHEP08(2018)166}{\emph{JHEP} {\bfseries 08}
  (2018) 166} [\href{https://arxiv.org/abs/1805.04377}{{\ttfamily
  1805.04377}}].

\bibitem{He:2017aed}
Y.-H. He, \emph{{Deep-Learning the Landscape}},
  \href{https://arxiv.org/abs/1706.02714}{{\ttfamily 1706.02714}}.

\bibitem{He:2017set}
Y.-H. He, \emph{{Machine-learning the string landscape}},
  \href{https://doi.org/10.1016/j.physletb.2017.10.024}{\emph{Phys. Lett. B}
  {\bfseries 774} (2017) 564}.

\bibitem{Krefl:2017yox}
D.~Krefl and R.-K. Seong, \emph{{Machine Learning of Calabi-Yau Volumes}},
  \href{https://doi.org/10.1103/PhysRevD.96.066014}{\emph{Phys. Rev. D}
  {\bfseries 96} (2017) 066014}
  [\href{https://arxiv.org/abs/1706.03346}{{\ttfamily 1706.03346}}].

\bibitem{Ruehle:2017mzq}
F.~Ruehle, \emph{{Evolving neural networks with genetic algorithms to study the
  String Landscape}},
  \href{https://doi.org/10.1007/JHEP08(2017)038}{\emph{JHEP} {\bfseries 08}
  (2017) 038} [\href{https://arxiv.org/abs/1706.07024}{{\ttfamily
  1706.07024}}].

\bibitem{Carifio:2017bov}
J.~Carifio, J.~Halverson, D.~Krioukov and B.~D. Nelson, \emph{{Machine Learning
  in the String Landscape}},
  \href{https://doi.org/10.1007/JHEP09(2017)157}{\emph{JHEP} {\bfseries 09}
  (2017) 157} [\href{https://arxiv.org/abs/1707.00655}{{\ttfamily
  1707.00655}}].

\bibitem{Erbin:2018csv}
H.~Erbin and S.~Krippendorf, \emph{{GANs for generating EFT models}},
  \href{https://arxiv.org/abs/1809.02612}{{\ttfamily 1809.02612}}.

\bibitem{Hashimoto:2018ftp}
K.~Hashimoto, S.~Sugishita, A.~Tanaka and A.~Tomiya, \emph{{Deep learning and
  the AdS/CFT correspondence}},
  \href{https://doi.org/10.1103/PhysRevD.98.046019}{\emph{Phys. Rev. D}
  {\bfseries 98} (2018) 046019}
  [\href{https://arxiv.org/abs/1802.08313}{{\ttfamily 1802.08313}}].

\bibitem{He:2018jtw}
Y.-H. He, \emph{{The Calabi-Yau Landscape: from Geometry, to Physics, to
  Machine-Learning}},  \href{https://arxiv.org/abs/1812.02893}{{\ttfamily
  1812.02893}}.

\bibitem{RUEHLE20201}
F.~Ruehle, \emph{Data science applications to string theory},
  \href{https://doi.org/https://doi.org/10.1016/j.physrep.2019.09.005}{\emph{Physics
  Reports} {\bfseries 839} (2020) 1 }.

\bibitem{Comsa:2019rcz}
I.~M. Comsa, M.~Firsching and T.~Fischbacher, \emph{{SO(8) Supergravity and the
  Magic of Machine Learning}},
  \href{https://doi.org/10.1007/JHEP08(2019)057}{\emph{JHEP} {\bfseries 08}
  (2019) 057} [\href{https://arxiv.org/abs/1906.00207}{{\ttfamily
  1906.00207}}].

\bibitem{Krishnan:2020sfg}
C.~Krishnan, V.~Mohan and S.~Ray, \emph{{Machine Learning ${\cal N}=8, D=5$
  Gauged Supergravity}},
  \href{https://doi.org/10.1002/prop.202000027}{\emph{Fortsch. Phys.}
  {\bfseries 68} (2020) 2000027}
  [\href{https://arxiv.org/abs/2002.12927}{{\ttfamily 2002.12927}}].

\bibitem{Chernodub:2019kon}
M.~Chernodub, H.~Erbin, I.~Grishmanovskii, V.~Goy and A.~Molochkov,
  \emph{{Casimir effect with machine learning}},
  \href{https://arxiv.org/abs/1911.07571}{{\ttfamily 1911.07571}}.

\bibitem{Chernodub:2020nip}
M.~Chernodub, H.~Erbin, V.~Goy and A.~Molochkov, \emph{{Topological defects and
  confinement with machine learning: the case of monopoles in compact
  electrodynamics}},  \href{https://arxiv.org/abs/2006.09113}{{\ttfamily
  2006.09113}}.

\bibitem{Betzler:2020rfg}
P.~Betzler and S.~Krippendorf, \emph{{Connecting Dualities and Machine
  Learning}}, \href{https://doi.org/10.1002/prop.202000022}{\emph{Fortsch.
  Phys.} {\bfseries 68} (2020) 2000022}
  [\href{https://arxiv.org/abs/2002.05169}{{\ttfamily 2002.05169}}].

\bibitem{Bao:2020nbi}
J.~Bao, S.~Franco, Y.-H. He, E.~Hirst, G.~Musiker and Y.~Xiao, \emph{{Quiver
  Mutations, Seiberg Duality and Machine Learning}},
  \href{https://arxiv.org/abs/2006.10783}{{\ttfamily 2006.10783}}.

\bibitem{He:2020eva}
Y.-H. He, E.~Hirst and T.~Peterken, \emph{{Machine-Learning Dessins d'Enfants:
  Explorations via Modular and Seiberg-Witten Curves}},
  \href{https://arxiv.org/abs/2004.05218}{{\ttfamily 2004.05218}}.

\bibitem{Simmons-Duffin:2015qma}
D.~Simmons-Duffin, \emph{{A Semidefinite Program Solver for the Conformal
  Bootstrap}}, \href{https://doi.org/10.1007/JHEP06(2015)174}{\emph{JHEP}
  {\bfseries 06} (2015) 174}
  [\href{https://arxiv.org/abs/1502.02033}{{\ttfamily 1502.02033}}].

\bibitem{Landry:2019qug}
W.~Landry and D.~Simmons-Duffin, \emph{{Scaling the semidefinite program solver
  SDPB}},  \href{https://arxiv.org/abs/1909.09745}{{\ttfamily 1909.09745}}.

\bibitem{Poland:2018epd}
D.~Poland, S.~Rychkov and A.~Vichi, \emph{{The Conformal Bootstrap: Theory,
  Numerical Techniques, and Applications}},
  \href{https://doi.org/10.1103/RevModPhys.91.015002}{\emph{Rev. Mod. Phys.}
  {\bfseries 91} (2019) 015002}
  [\href{https://arxiv.org/abs/1805.04405}{{\ttfamily 1805.04405}}].

\bibitem{Ashmore:2019wzb}
A.~Ashmore, Y.-H. He and B.~A. Ovrut, \emph{{Machine learning Calabi-Yau
  metrics}},  \href{https://arxiv.org/abs/1910.08605}{{\ttfamily 1910.08605}}.

\bibitem{MEHTA20191}
P.~Mehta, M.~Bukov, C.-H. Wang, A.~G. Day, C.~Richardson, C.~K. Fisher et~al.,
  \emph{A high-bias, low-variance introduction to machine learning for
  physicists},
  \href{https://doi.org/https://doi.org/10.1016/j.physrep.2019.03.001}{\emph{Physics
  Reports} {\bfseries 810} (2019) 1 }.

\bibitem{Bao:2020sqg}
J.~Bao, Y.-H. He, E.~Hirst and S.~Pietromonaco, \emph{{Lectures on the
  Calabi-Yau Landscape}},  \href{https://arxiv.org/abs/2001.01212}{{\ttfamily
  2001.01212}}.

\bibitem{sirignano2018dgm}
J.~Sirignano and K.~Spiliopoulos, \emph{Dgm: A deep learning algorithm for
  solving partial differential equations}, {\emph{Journal of Computational
  Physics} {\bfseries 375} (2018) 1339}.

\bibitem{Piscopo:2019txs}
M.~L. Piscopo, M.~Spannowsky and P.~Waite, \emph{{Solving differential
  equations with neural networks: Applications to the calculation of
  cosmological phase transitions}},
  \href{https://doi.org/10.1103/PhysRevD.100.016002}{\emph{Phys. Rev. D}
  {\bfseries 100} (2019) 016002}
  [\href{https://arxiv.org/abs/1902.05563}{{\ttfamily 1902.05563}}].

\bibitem{broecker2017machine}
P.~Broecker, J.~Carrasquilla, R.~G. Melko and S.~Trebst, \emph{Machine learning
  quantum phases of matter beyond the fermion sign problem}, {\emph{Scientific
  reports} {\bfseries 7} (2017) 1}.

\bibitem{Krippendorf:2020gny}
S.~Krippendorf and M.~Syvaeri, \emph{{Detecting Symmetries with Neural
  Networks}},  \href{https://arxiv.org/abs/2003.13679}{{\ttfamily 2003.13679}}.

\bibitem{726791}
Y.~{Lecun}, L.~{Bottou}, Y.~{Bengio} and P.~{Haffner}, \emph{Gradient-based
  learning applied to document recognition}, {\emph{Proceedings of the IEEE}
  {\bfseries 86} (1998) 2278}.

\bibitem{10.5555/2999134.2999257}
A.~Krizhevsky, I.~Sutskever and G.~E. Hinton, \emph{Imagenet classification
  with deep convolutional neural networks},  in \emph{Proceedings of the 25th
  International Conference on Neural Information Processing Systems - Volume
  1}, NIPS’12, (Red Hook, NY, USA), p.~1097–1105, Curran Associates Inc.,
  2012.

\bibitem{Alessandretti:2019jbs}
L.~Alessandretti, A.~Baronchelli and Y.-H. He, \emph{{Machine Learning meets
  Number Theory: The Data Science of Birch-Swinnerton-Dyer}},
  \href{https://arxiv.org/abs/1911.02008}{{\ttfamily 1911.02008}}.

\bibitem{He:2019nzx}
Y.-H. He and M.~Kim, \emph{{Learning Algebraic Structures: Preliminary
  Investigations}},  \href{https://arxiv.org/abs/1905.02263}{{\ttfamily
  1905.02263}}.

\bibitem{Nakayama:2013is}
Y.~Nakayama, \emph{{Scale invariance vs conformal invariance}},
  \href{https://doi.org/10.1016/j.physrep.2014.12.003}{\emph{Phys. Rept.}
  {\bfseries 569} (2015) 1} [\href{https://arxiv.org/abs/1302.0884}{{\ttfamily
  1302.0884}}].

\bibitem{Hogervorst:2013kva}
M.~Hogervorst, H.~Osborn and S.~Rychkov, \emph{{Diagonal Limit for Conformal
  Blocks in $d$ Dimensions}},
  \href{https://doi.org/10.1007/JHEP08(2013)014}{\emph{JHEP} {\bfseries 08}
  (2013) 014} [\href{https://arxiv.org/abs/1305.1321}{{\ttfamily 1305.1321}}].

\bibitem{Chen:2019gka}
H.-Y. Chen and H.~Kyono, \emph{{On conformal blocks, crossing kernels and
  multi-variable hypergeometric functions}},
  \href{https://doi.org/10.1007/JHEP10(2019)149}{\emph{JHEP} {\bfseries 10}
  (2019) 149} [\href{https://arxiv.org/abs/1906.03135}{{\ttfamily
  1906.03135}}].

\bibitem{Cybenko1989}
G.~Cybenko, \emph{Approximation by superpositions of a sigmoidal function},
  \href{https://doi.org/10.1007/BF02551274}{\emph{Mathematics of Control,
  Signals and Systems} {\bfseries 2} (1989) 303}.

\bibitem{HORNIK1991251}
K.~Hornik, \emph{Approximation capabilities of multilayer feedforward
  networks},
  \href{https://doi.org/https://doi.org/10.1016/0893-6080(91)90009-T}{\emph{Neural
  Networks} {\bfseries 4} (1991) 251 }.

\bibitem{nielsenneural}
M.~A. Nielsen, \emph{Neural networks and deep learning},  2018.

\bibitem{59192}
A.~Geron, \emph{Hands-on machine learning with Scikit-Learn and TensorFlow :}.
  O'Reilly,, Beijing, first edition.~ed., 2017.

\bibitem{lecun-mnisthandwrittendigit-2010}
Y.~LeCun and C.~Cortes, \emph{{MNIST} handwritten digit database}, .

\bibitem{carrasquilla2017machine}
J.~Carrasquilla and R.~G. Melko, \emph{Machine learning phases of matter},
  {\emph{Nature Physics} {\bfseries 13} (2017) 431}.

\bibitem{ch2017machine}
K.~Ch’Ng, J.~Carrasquilla, R.~G. Melko and E.~Khatami, \emph{Machine learning
  phases of strongly correlated fermions}, {\emph{Physical Review X} {\bfseries
  7} (2017) 031038}.

\bibitem{JMLR:v18:17-527}
A.~Morningstar and R.~G. Melko, \emph{Deep learning the ising model near
  criticality}, {\emph{Journal of Machine Learning Research} {\bfseries 18}
  (2018) 1}.

\bibitem{d2020learning}
F.~D'Angelo and L.~B{\"o}ttcher, \emph{Learning the ising model with generative
  neural networks}, {\emph{Physical Review Research} {\bfseries 2} (2020)
  023266}.

\bibitem{Kos:2014bka}
F.~Kos, D.~Poland and D.~Simmons-Duffin, \emph{{Bootstrapping Mixed Correlators
  in the 3D Ising Model}},
  \href{https://doi.org/10.1007/JHEP11(2014)109}{\emph{JHEP} {\bfseries 11}
  (2014) 109} [\href{https://arxiv.org/abs/1406.4858}{{\ttfamily 1406.4858}}].

\bibitem{Fitzpatrick:2012yx}
A.~Fitzpatrick, J.~Kaplan, D.~Poland and D.~Simmons-Duffin, \emph{{The Analytic
  Bootstrap and AdS Superhorizon Locality}},
  \href{https://doi.org/10.1007/JHEP12(2013)004}{\emph{JHEP} {\bfseries 12}
  (2013) 004} [\href{https://arxiv.org/abs/1212.3616}{{\ttfamily 1212.3616}}].

\bibitem{Komargodski:2012ek}
Z.~Komargodski and A.~Zhiboedov, \emph{{Convexity and Liberation at Large
  Spin}}, \href{https://doi.org/10.1007/JHEP11(2013)140}{\emph{JHEP} {\bfseries
  11} (2013) 140} [\href{https://arxiv.org/abs/1212.4103}{{\ttfamily
  1212.4103}}].

\end{thebibliography}\endgroup

\end{document}